\documentclass[a4paper,australian,amsmath,aps,prd,twocolumn,noshowpacs]{revtex4-1}

\usepackage{times}
\usepackage{amsmath}
\usepackage{amsfonts}
\usepackage{amssymb}
\usepackage{graphicx}
\usepackage{mathrsfs}

\usepackage{xcolor}

\newcommand{\CSSM}{Special Research Centre for the Subatomic Structure
  of Matter (CSSM),\\School of Chemistry and Physics, University of
  Adelaide, South Australia 5005, Australia} 

\begin{document}

\title{Instanton contributions to the low-lying hadron mass spectrum}
\author{Samuel D. Thomas}
\author{Waseem Kamleh}
\author{Derek B. Leinweber}
\affiliation{\CSSM}

\date{\today}

\begin{abstract}
The role of instanton-like objects in the QCD vacuum on the mass
spectrum of low-lying light hadrons is explored in lattice QCD.  Using
over-improved stout-link smearing, tuned to preserve instanton-like
objects in the QCD vacuum, the evolution of the mass spectrum under
smearing is examined.  The calculation is performed using a
$20^3\times40$ dynamical fat-link-irrelevant-clover (FLIC) fermion
action ensemble with lattice spacing 0.126 fm. Through the consideration of a range of pion masses, the effect of the vacuum instanton content is compared at a common pion mass.  While the
qualitative features of ground-state hadrons are preserved on
instanton-dominated configurations, the excitation spectrum
experiences significant changes.  The underlying physics revealed
shows little similarity to the direct-instanton interaction
predictions of the instanton liquid model.
\end{abstract}

\maketitle

\section{Introduction}
The instanton \cite{Belavin197585} is well known as a classical
solution to the pure-gauge Yang-Mills equations.  It has topological
change $\pm 1$, action $8\pi^2 /g^2$, and is associated with a
localised zero eigenmode of the Dirac operator. Various models exist
for the QCD vacuum as composed purely of superpositions of these
objects and here we consider the random instanton liquid model (RILM) \cite{Shuryak-A1,Shuryak-A3,Shuryak-B,Shuryak-C1,Shuryak-C3,Horvath:2002gk} as a point of comparison.  Phenomenology constrains the RILM model parameters including $\rho_{\rm inst} \approx$ 0.33~fm for the average instanton size and $n \approx 1 ~\textrm{fm}^{-4}$ for the pseudo-particle density.

Such a model vacuum generates approximate Dirac zero-modes with
definite chirality.  A quark propagating via a zero-mode changes its
chirality.  Thus, the masses of the pseudoscalar channels (the pion,
and the diquark in the nucleon) have direct instanton-induced
contributions which are attractive and reduce their mass.  By
comparison, the rho as a vector meson has instanton contributions only
at a higher level (analogous to a 2-$\pi$
intermediate state). Similarly, the Delta has interactions only at the 
6-pseudoparticle level and higher \cite{Shuryak-prop-1,Shuryak-prop-2,Shuryak-prop-3,schaefer-1998,Shuryak:2014gja}.

We wish to determine what role the instantons present in the QCD
vacuum of lattice QCD simulations play in the determination of hadron
phenomenology.  Starting from a Monte-Carlo generated calculation of
QCD on the lattice, we filter out the short-distance gluonic
interactions such that the underlying instanton degrees of freedom are
revealed.  We use over-improved stout-link smearing
\cite{Moran:2008ra} to do this and the merits of this approach
are discussed in Sec.~\ref{sec:revealing}.
 
Section \ref{sec:methods} provides an overview of the lattice QCD
simulation methods and associated parameters.  It also describes our
correlation matrix approach which enables us to accurately determine
both the ground-state hadron spectrum and the first radial
excitations.

We will then examine the low-lying hadron spectrum as a function of
smearing in Sec.~\ref{sec:results}, monitoring its evolution as the
QCD vacuum progresses from having significant topological structure,
most of it not instanton-like, through to being both
instanton-dominated and eventually sparse as nearby
instanton--anti-instanton pairs annihilate.  Conclusions are drawn in
Sec.~\ref{sec:conclusion}.

\section{Revealing Instantons}
\label{sec:revealing}

Early methods of smoothing short-distance fluctuations to reveal the
underlying instanton degrees of freedom used cooling.  This proceeds
by replacing each link $U_\mu(x)$ on the lattice by a new link such
that the gluonic action is minimised.  Unfortunately, this approach
also tends to remove the topological configurations of interest from
the lattice. This occurs due to the discretisation error involved in
minimizing the local action.  Expanding the gluonic Wilson action in
terms of a single-instanton solution one finds \cite{perez1994}
\begin{equation}
S_{\textrm{inst}} = \frac{8\pi^2}{g^2} \left\lbrace 1 - 
\frac{1}{5}\left(\frac{a}{\rho}\right)^2 + O(a^4)\right\rbrace \, .
\end{equation}
Thus the discretisation error enables one to reduce the action by
reducing the instanton size parameter $\rho$.  Upon minimizing the
action, instantons shrink.  At sufficient cooling, they will become
small enough that significant discretisation errors will allow them to
`fall through' the lattice.

It is possible, however, to include larger combinations of loops
having different discretisation errors. Coefficients can be chosen
such that the $O(a^2)$ error is cancelled \cite{symanzik}, giving the
`improved' action. Unfortunately this method still leads to a negative
leading $O(a^4)$ discretisation error, and the same unwanted corrosion
of topological objects. Higher order terms composed from combinations
of larger loops can also be included in the action, but this requires
increasing accuracy on the perturbative corrections to the improvement
coefficients, typically estimated via the mean link of tadpole
improvement.

Instead, one can adopt the approach of over-improvement
\cite{perez1994,Moran:2008ra} and express the action in terms of
an improvement parameter $\epsilon$,
\begin{align}
S(\epsilon) &= \beta\sum_x\sum_{\mu > \nu} \left \{
\frac{5-2\epsilon}{3} \left ( 1-P_{\mu\nu}(x)\, \right ) \right
. \nonumber \\ 
&{} \qquad \left . - \frac{1 - \epsilon}{12} \left [
\left (1-R_{\mu\nu}(x)\, \right ) + \left (1 - R_{\nu\mu}(x)\, \right )
\right ] \right \} \, ,
\label{eq:over-imp}
\end{align}
where $P_{\mu\nu}(x)$ denotes 1/3 of the real trace of the clover
average of the four plaquettes touching the point $x$ and similarly
$R_{\mu\nu}(x)$ denotes denotes 1/3 of the real trace of the clover
average of four $2\times 1$ Wilson loops. The choice of $2\times 1$
rectangles over the $2\times 2$ squares in
Ref.~\cite{Moran:2008ra} is in the interest of preserving
locality and minimizing the number of links. The coefficients are
chosen such that $\epsilon=1$ gives the unimproved Wilson action and
$\epsilon=0$ provides an $O(a^2)$-improved action.
We can also compare the behavior of the smearing algorithm to that provided by the Wilson or gradient flow, as in Ref.~\cite{Bonati:2014tqa}. For $\epsilon=0$, $N$ sweeps of stout-link smearing is equivalent to $t=\alpha N$ for the Wilson flow time as long as the smearing parameter $\alpha$ is sufficiently small. We use $\alpha=0.06$ which is even smaller than the standard $\alpha=0.1$ for stout-link smearing.

A negative value for the $\epsilon$ parameter will lead to a positive
leading-order discretisation error which inhibits the shrinking of
instanton-like structures under smearing.  However a large negative
value would cause instantons to grow under smearing.  We use an
$\epsilon$ value of $-0.25$, as recommended in Ref.~\cite{Moran:2008ra} providing the required stability with
marginal discretisation error. The effect of $\epsilon$ is also studied in Sec.~\ref{sec:results}.

This scheme preserves instanton-like objects with a size parameter
$\rho$ greater than the dislocation threshold of $1.97\, a$
\cite{Moran:2008ra}.  Herein, $a = 0.126$ fm such that
instantons of size $\rho < 0.25$ fm will be removed under
over-improved stout-link smearing \cite{Moran:2008ra}.  However,
this effect may be regarded as small.  The scale dependence of the
instanton action $S_0 = {8\pi^2}/{g^2}$ on the coupling
constant $g$ in the context of asymptotic freedom suppresses the
presence of small instantons.  A study of the instanton distribution
within dynamical gauge fields with light dynamical quarks provides
\cite{Moran:2008qd} $\rho_{\rm inst} = 0.415$ fm with the standard
deviation of the distribution of instanton sizes of only 0.075 fm
indicating a sharply peaked distribution with few small-size
instantons.

An additional effect of smearing is that any smearing algorithm
designed to suppress short-distance perturbative interactions from the
gauge field will also tend to annihilate closely spaced
instanton--anti-instanton pairs.  As the effect of this is mainly to
reduce the pseudo-particle density, we expect a corresponding change in
the quark condensate \cite{Shuryak-B}.  We will use this behaviour of
the smearing algorithm to examine the lattice QCD vacuum as it
progresses from having significant topological structure, most of it
not instanton-like, through to being both instanton-dominated and
sparse.

\section{Simulation Methods}
\label{sec:methods}

\subsection{Correlation Functions and the Variational Approach}
\label{subsec:corrFun}

The low-lying hadronic masses can be extracted from the lattice by
analysis of their corresponding 2-point correlation functions. This
correlation function is defined as $G_2(x) = \langle 0 \vert
\chi(x)\bar{\chi}(0)\vert 0\rangle$, where $\chi$ is the interpolating
field corresponding to the hadron of interest.  The correlation
function, in momentum space, is then of the form 
\begin{equation}
G_2(\vec p, t) = \sum_{\vec x}
\exp(-i\vec p \cdot \vec x)\, \langle\, 0\, |\, \chi(x)\, \bar{\chi}(0)\, |
\, 0 \,\rangle \, .
\end{equation}
Inserting a complete set of states, $| B \rangle$, and utilising
the translation operator, one obtains the Euclidean time correlator
\begin{equation}
G_2(\vec{p},t) = \sum_B e^{-E_{B}(\vec{p})\, t} \lambda_B(\vec{p})\, \bar{\lambda}_B(\vec{p})
\end{equation}
with $\lambda_B(\vec{p}) = \langle\, 0 \, | \, \chi \, | \, B, \vec p
\, \rangle $ describing the coupling of the state $B$ with momentum
$\vec{p}$ to the operator $\chi$ and $E_{B}(\vec{p})$ the on-shell
energy of the state.

The determination of the mass of the lowest lying state is hampered by
the tower of exponentials from excited state contributions.  Although
these are suppressed by the factor $e^{-E_B t}$, it is often difficult
to wait until a sufficiently large Euclidean time where all excited
state contaminations have vanished.  Moreover, with broad fermion
source smearings and narrow fermion sink smearings, excited states can
enter with a negative weight and create false plateaus.

This problem is particularly challenging on smeared configurations.
As we will see, hadronic excited states do not maintain a significant
mass splitting from the ground-state hadrons making the extraction of
even the ground-state mass difficult.

The solution to the state isolation problem is now well established.
One considers a matrix of correlation functions in a variational
analysis \cite{Michael:1985ne,Luscher:1990ck,McNeile:2000xx}.  The
operators used to create the correlation functions can be chosen to
have any form (as long as they have the correct quantum numbers).
Different operators have different couplings $\lambda_B$ to each
state, $B$ and one seeks linear combinations of the operators
constructed to isolate each state of the spectrum.

The best approach for isolating a state within a tower of states
excited by a particular interpolating field is to introduce
differently sized covariantly smeared sources and sinks
\cite{Mahbub:2010rm,Mahbub:2013ala,Owen:2012ts}.  Physical hadrons are
extended objects, and a linear combination of Gaussian sources allows
one to approximate the hadronic wave function \cite{Roberts:2013oea}.
This allows for a precise determination of ground-state properties.
The suppression of excited states provides ground-state isolation
early in Euclidean time.  The approach also provides access to the
excited states having the same spin and parity.

We solve the generalized eigenvalue problem for the matrix $G_{ij}$
whose elements are the correlation functions generated from the
operators $\chi_i$ and $\bar{\chi}_j$, normalised at the fermionic
source time.  If $\bar\phi^\alpha = u^\alpha_j\, \bar\chi_j$ is an
operator constructed to isolate state $\alpha$, then the recurrence
relation relating times $t_0$ and $t_0+dt$
\begin{equation}
G_{ij}(t_0+dt)\, u^\alpha_j = e^{-E_\alpha dt}\, G_{ij}(t_0) \,
u^\alpha_j \, ,
\end{equation}
can be used to construct a generalised eigenvalue equation for the
right eigenvector $u^\alpha$ and eigenvalue $e^{-E_\alpha\, dt}$
\begin{equation}
G^{-1}(t_0)\, G(t_0+dt)\, u^\alpha = e^{-E_\alpha\, dt}\, u^\alpha
\, .
\end{equation}
The reference times $t_0$ and interval $dt$ must be selected to lie
within a region where the excited state contributions are strong and
not yet exponentially suppressed.  However, some amount of Euclidean
time evolution is helpful in reducing the number of states
contributing significantly in the correlators to the dimension of the
correlation matrix such that state isolation is achieved.

While the eigenvalues of the generalised eigenvalue equations depend
strongly on the values of these variational parameters, $t_0$ and
$dt$, the associated eigenvectors are robust against this variation.
A similar analysis can be done to obtain the left eigenvector
$v^\alpha$ from which a state-projected correlator $v^\alpha_i\,
G_{ij}(t)\, u^\alpha_j$ can be constructed.  

This correlator is insensitive to the variational parameters.  In
practice, a $t_0$ value one or two time steps from the source
accompanied by a $dt$ value of two or three lattice time steps
provides good eigenstate isolation in the projected correlator.  One
can then apply standard analysis techniques using the covariance
matrix-based $\chi^2$ per degree of freedom to carefully identify the
Euclidean time regime dominated by a single eigenstate.

\subsection{Simulation Methods and Parameters}

Previous works in this vein \cite{Chu:1994vi, schaefer-1994, Negele:1997na, DeGrand199897} were considered in the mid to late 1990s.  While they often considered the quenched approximation, a greater concern is the
use of standard cooling algorithms based on the Wilson gauge action.

As discussed in Sec.~\ref{sec:revealing}, this algorithm rapidly
destroys the instanton-like topoological structures that one is
attempting to study.  In this case, the final smeared configuration is
much closer to a dilute instanton gas than to an instanton liquid.

The calculations presented herein are performed on an ensemble of 76 2-flavour
$20^3 \times 40$ gauge field configurations generated using dynamical fat-link-irrelevant-clover (FLIC) fermions
\cite{Zanotti:2001yb,Zanotti:2004dr,Boinepalli:2004fz,Kamleh:2004xk,Zanotti:2004qn},
with lattice coupling $\beta=3.94$ and an $SU(3)$-flavour symmetric hopping parameter $\kappa = 0.1324$, providing $m_\pi = 540~\rm{MeV}$.  The lattice spacing associated with the string tension is $a = 0.126$ fm, providing a spatial extent of 2.52 fm. As a smearing sweep is a local short-distance effect that does not affect the string tension, we consider the lattice spacing to be unaltered under a smearing sweep.  Cumulative smearing sweeps do affect long-distance physics and we will examine this effect in the following section.

The smearing of the gauge field is performed using over-improved stout-link smearing \cite{Moran:2008ra} with $\epsilon=-0.25$ and an isotropic smearing parameter $\alpha_{sm}=0.06$, smaller than the typical value \cite{Morningstar:2003gk} of 0.10.

Valence quark propagators are calculated via the FLIC fermion action at multiple $\kappa$ values.  The boundary conditions are periodic in the spatial dimensions, and fixed in the Euclidean time dimension. The fermionic source is inserted away from the boundary at $t=10$, sufficient to avoid artifacts associated with the boundary. In constructing a basis for our variational approach to isolating states, we use gauge invariant Gaussian smeared \cite{Gusken:1989qx} fermion sources.  The smearing procedure is: 
\begin{align}
\psi_{i}(x,t) &=\sum_{x'}\, F(x,x')\, \psi_{i-1}(x',t) \, ,
\end{align}
where
\begin{align}
F(x,x') &= {(1-\alpha)}\, \delta_{x,x'} \\
        &+ \frac{\alpha}{6}
\sum_{\mu=1}^{3} \left [ U_{\mu}(x)\, \delta_{x',x+\hat\mu} 
                + U_{\mu}^{\dagger}(x-\hat\mu)\,
                \delta_{x',x-\hat\mu} \right ],
\nonumber
\end{align}
and the parameter $\alpha=0.7$ is used in our calculation. After repeating the procedure $N_{\rm sm}$ times on a point source the resulting smeared fermion field is,
\begin{align}
\psi_{N_{\rm sm}}(x,t) &=\sum_{x'}F^{N_{\rm sm}}(x,x')\psi_{0}(x',t).
\end{align}
We consider $N_{\rm sm}$ = 10, 25, 50, 100, and 150 sweeps in constructing effective correlation-matrix bases.

\section{Results}
\label{sec:results}

\subsection{Gluonic observables}

\begin{figure}
\begin{center}
\includegraphics[width=0.95\hsize]{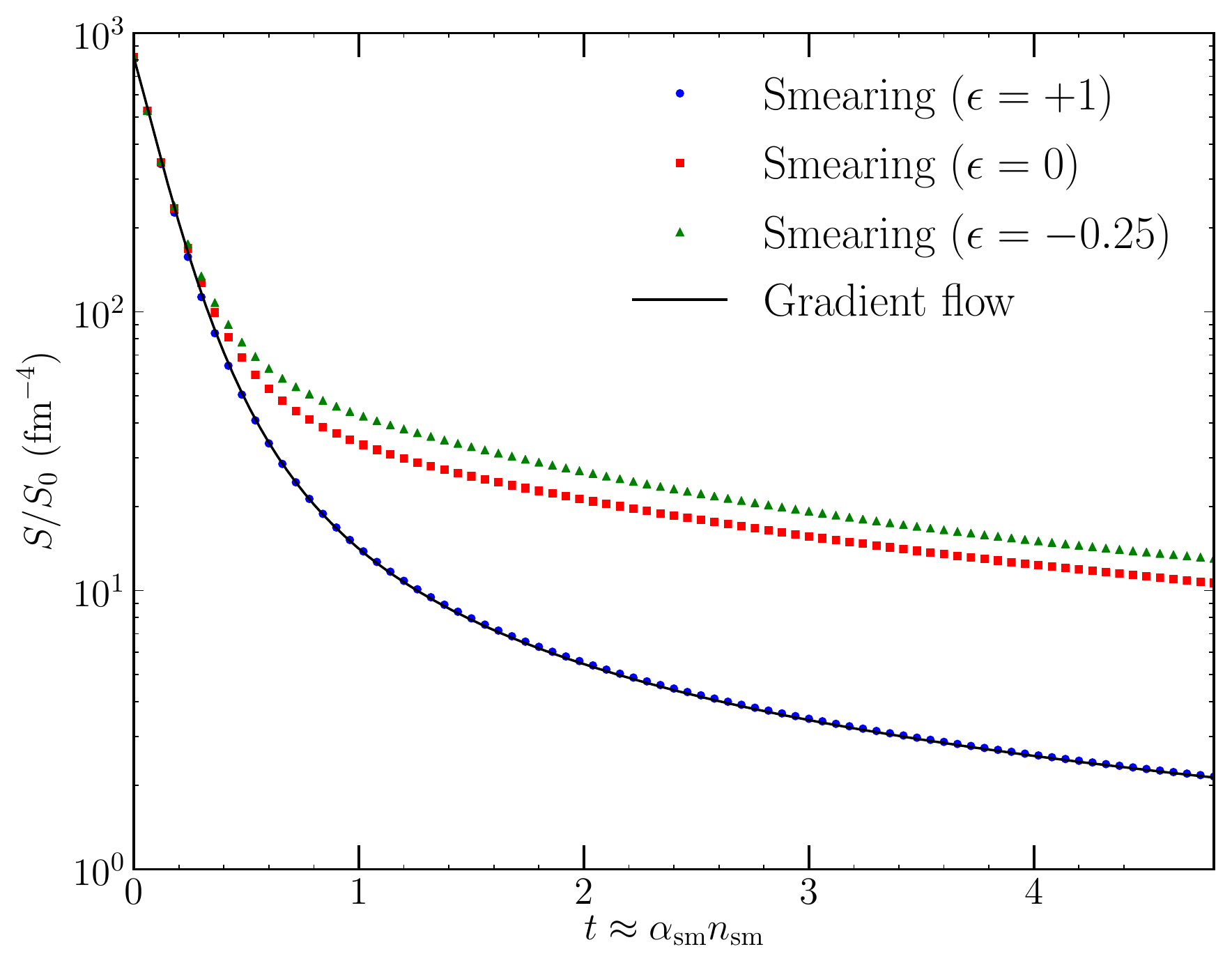}
\end{center}
\caption{The evolution of the action density for a typical configuration under various methods of smearing. $\epsilon=1$ is the Wilson action, and $\epsilon=0$ corresponds to the $\mathcal{O}(a^2)$-improved action. The action under over-improved smearing ($\epsilon=-0.25$) lies above either of these as required. With the identification $t=\alpha_{\rm{sm}}n_{\rm{sm}}$, numerical integration of the Wilson flow agrees with smearing using the Wilson action. The highest level of smearing considered in the following sections (60 sweeps) is equivalent to a Wilson flow time of $t=3.6$.}
\label{fig:gluon_S}
\end{figure}

We begin by examining the effect of stout-link smearing on the action of our gauge-field ensemble.  Figure~\ref{fig:gluon_S} displays the evolution of the action as a function of the number of smearing sweeps.  For our choice of $\epsilon=-0.25$, after only 10 sweeps of smearing, the action has dropped to one tenth of its initial value. The majority of the short-distance interactions are removed within the first few iterations of smearing.  It takes another 50 sweeps of smearing for the action to reduce by another order of magnitude. The Wilson flow is also shown for comparison of smearing extent. We will focus on ensembles following 10, 20, 40, and 60 sweeps of smearing; these correspond to Wilson flow times of 0.6, 1.2, 2.4, and 3.6 respectively.

\begin{figure}
\begin{center}
\includegraphics[clip=true,width=0.95\hsize]{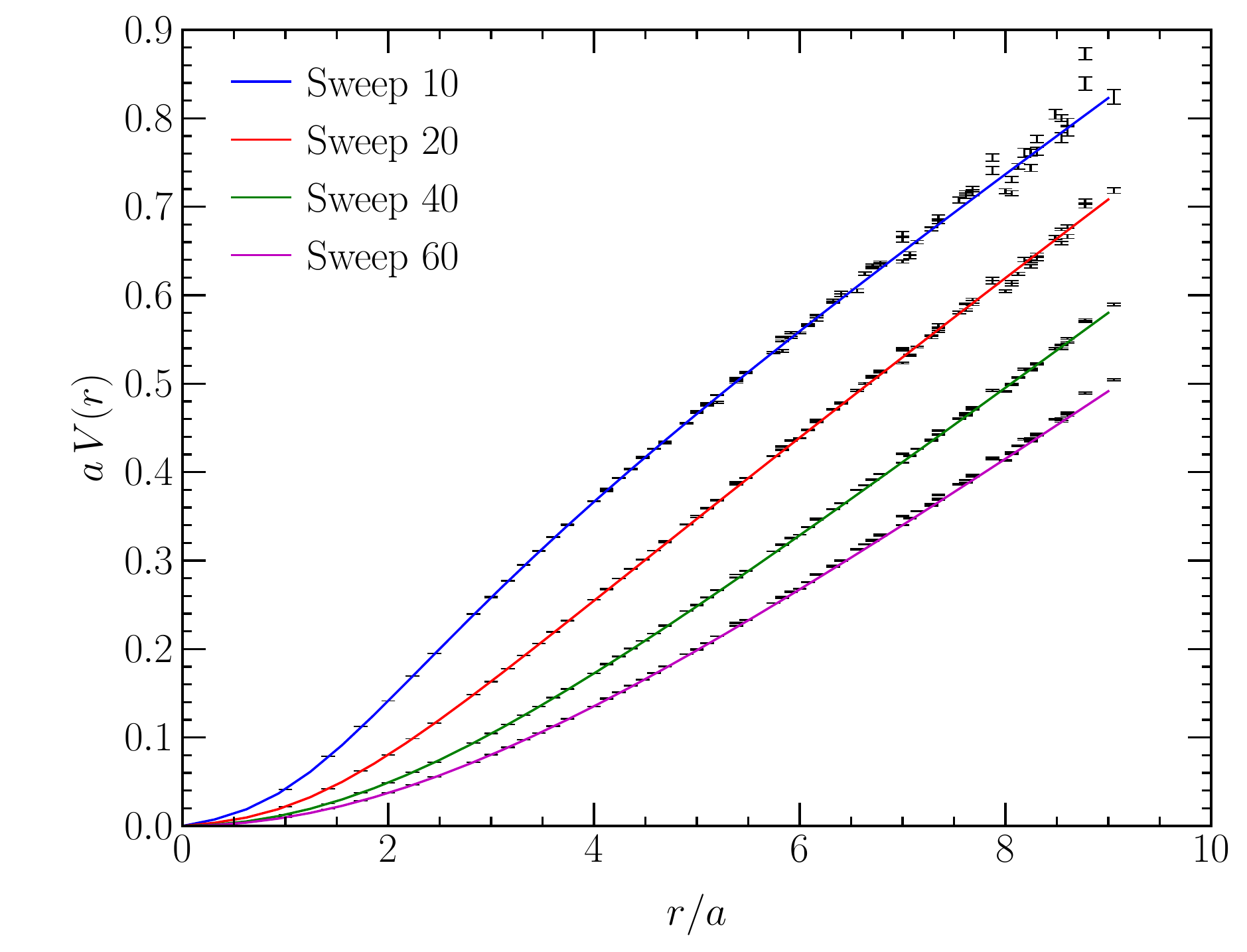}
\end{center}
\caption{The evolution of the static quark potential under various
  levels of over-improved stout-link smearing.}
\label{fig:sqp}
\end{figure}

In Fig.~\ref{fig:sqp}, the static quark potential, determined from the Wilson loop, is shown for configurations at various levels of smearing.  The attractive Coulomb-potential behavior at small $r$ is lost rapidly under smearing as the algorithm removes short-distance interactions.  The fit is a third-order Pad\'e approximant constrained by the short-distance behaviour $V \rightarrow V_{free}=0$ and the long-distance behaviour of a linearly rising potential.  It has the form
\begin{equation}
V(r) = \frac{a_1\, r + a_2\, r^2 + a_3\, r^3}{b_0 + b_1\, r + r^2} \, .
\label{eq:pade}
\end{equation}
The large $r$ slope of the potential reflects the approximate invariance of the string tension and associated lattice spacing. Best fit parameters are summarised in Table~\ref{tab:padeParam}.

%

\begin{table}[t]
\caption{The best fit parameters for the Pad\'e of Eq.~(\ref{eq:pade})
  fit to the lattice QCD results for the static quark potential
  illustrated in Fig.~\ref{fig:sqp}.  Uncertainties in the parameters
  are determined from a single-elimination jackknife analysis.  }
\label{tab:padeParam}
\begin{ruledtabular}
\begin{tabular}{llllll}
Sweep & $a_1$ & $a_2$ & $a_3$ & $b_0$ & $b_1$              \\ 
\hline                                                                       
10 & 0.093(2)  & 0.01(1)  & 0.082(1) & 4.7(2)  & -1.22(5)  \\ 
20 & 0.062(1)  & 0.020(4) & 0.084(1) & 6.4(2)  &  0.22(3)  \\ 
40 & 0.035(1)  & 0.212(4) & 0.081(1) & 21.4(3) &  3.28(5)  \\ 
60 & 0.02(5)   & 0.384(1) & 0.073(1) & 42(1)   &  5(1)     \\ 
\end{tabular}
\end{ruledtabular}
\end{table}

%

For sufficiently large $r$, the parameter $a_3$ is equivalent to the
string tension $\sigma$ and we note it remains almost constant under
moderate levels of smearing.  This differs significantly from the
behavior reported in Ref.~\cite{Chu:1994vi} where the string tension
was reduced to only 27\% of its original value after 25 sweeps of the
unimproved cooling algorithm.  This also differs from the theoretical
prediction of the instanton liquid model, which was shown to generate
a string tension \cite{Brower1999512}, but with a value much smaller
than the physical value.  The over-improved stout-link smearing
algorithm tuned specifically to preserve instanton-like structure in
the gauge field configurations retains the long-distance string
tension remarkably well.

The process of stout-link smearing is expected to remove the short
distance physics up to an effective radius \cite{Bakry:2011ew} which
may be parameterised as
\begin{equation}
R = a\, \left ( c\, \rho_{sm}\, N_{\rm sw} \right )^{1/2} \, ,
\end{equation}
under the random-walk hypothesis.  The coefficient $c$ is a
proportionality constant determined \cite{Bakry:2011ew} to be $c =
6.15(3)$.  The smearing radii for our selected values of $N_{\rm sw} =
0$, 10, 20, 40, 60 are $R/a = 0$, 1.9, 2.7, 3.8, and 4.7 respectively.
While this explains the preservation of the string tension observed in
Fig.~\ref{fig:sqp} and Table~\ref{tab:padeParam}, one needs to examine
the extent to which an ensemble of instantons has been isolated in the
QCD vacuum.

The instanton content of the vacuum under the same over-improved
stout-link smearing algorithm selected herein was also studied in
Ref.~\cite{Trewartha:2013qga}.  To examine the extent to which the
non-trivial topology identified on the lattice is consistent with
instantons, two measures of the local maxima of the action density
found on representative configurations were measured and compared to
the classical instanton solution.  The instanton size is measured by
fitting the profile of the action density in a $(2 a)^4$ hypercube
surrounding the position of the local maximum to the classical
instanton action density
\begin{equation}
S_{0}(x) = \xi 
\frac{6}{\pi^{2}}\frac{\rho^{4}}{((x-x_{0})^{2}+\rho^{2})^{4}} \, .
\label{eq:instactdens}
\end{equation}
Here $\xi$, $\rho$ and $x_{0}$ are fit parameters, noting that $x_{0}$
is not restricted to a lattice site. The parameter $\xi$ is introduced
as lattice topological objects often have a higher action than
classical instantons.  We wish to determine the size, $\rho$, by using
the shape of the action density around the local maximum, rather than
the height. Considering the relative RMS deviation over the $3^4$ hypercube of points surrounding the local maximum, $\sqrt{ \frac{1}{V-1} \sum_{x\in V} ( S_0(x) - S(x) )^2 }/S_0$,
we find (Eq.~\ref{eq:instactdens}) fits the data with a typical percentage deviation of $10\% $. This deviation decreases as the gauge fields are smeared.

We can then compare the size of instanton candidates to
the value of the topological charge at the centre of an
(anti-) instanton in the context of the classical instanton
relationship
\begin{equation}
\label{topcharge}
q(x_{0})=Q\frac{6}{\pi^{2} \rho^{4}} \, ,
\end{equation}
where $Q=\mp 1$ for an (anti-)instanton.  $q(x_{0})$ at the fitted
values of $x_{0}$ are found using linear interpolation from
neighbouring hypercubes to find an extremum inside the hypercube
containing $x_{0}$.

\begin{figure*}[t]
\begin{center}
\includegraphics[ clip=true,width=0.45\hsize]{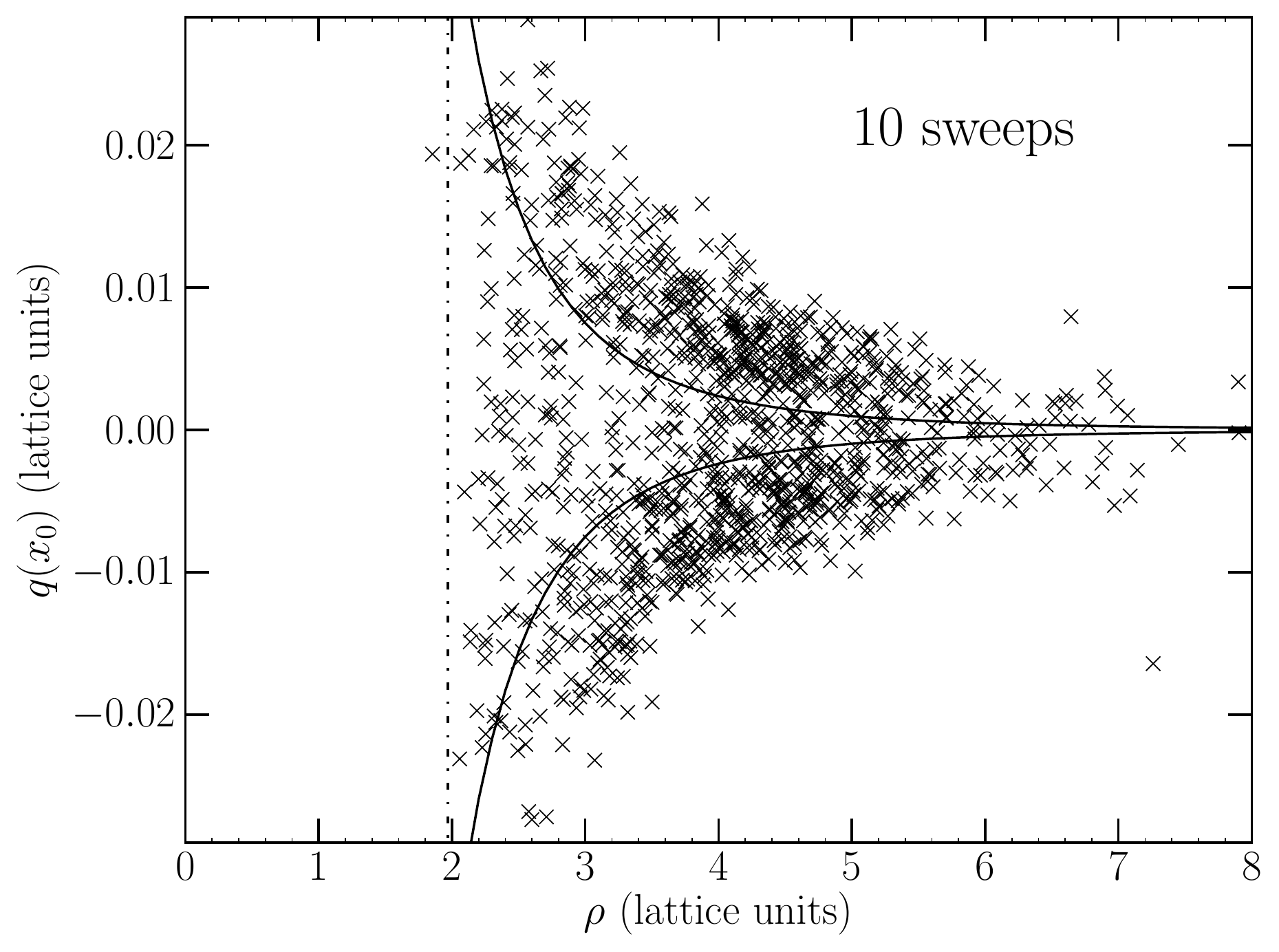}\quad
\includegraphics[ clip=true,width=0.45\hsize]{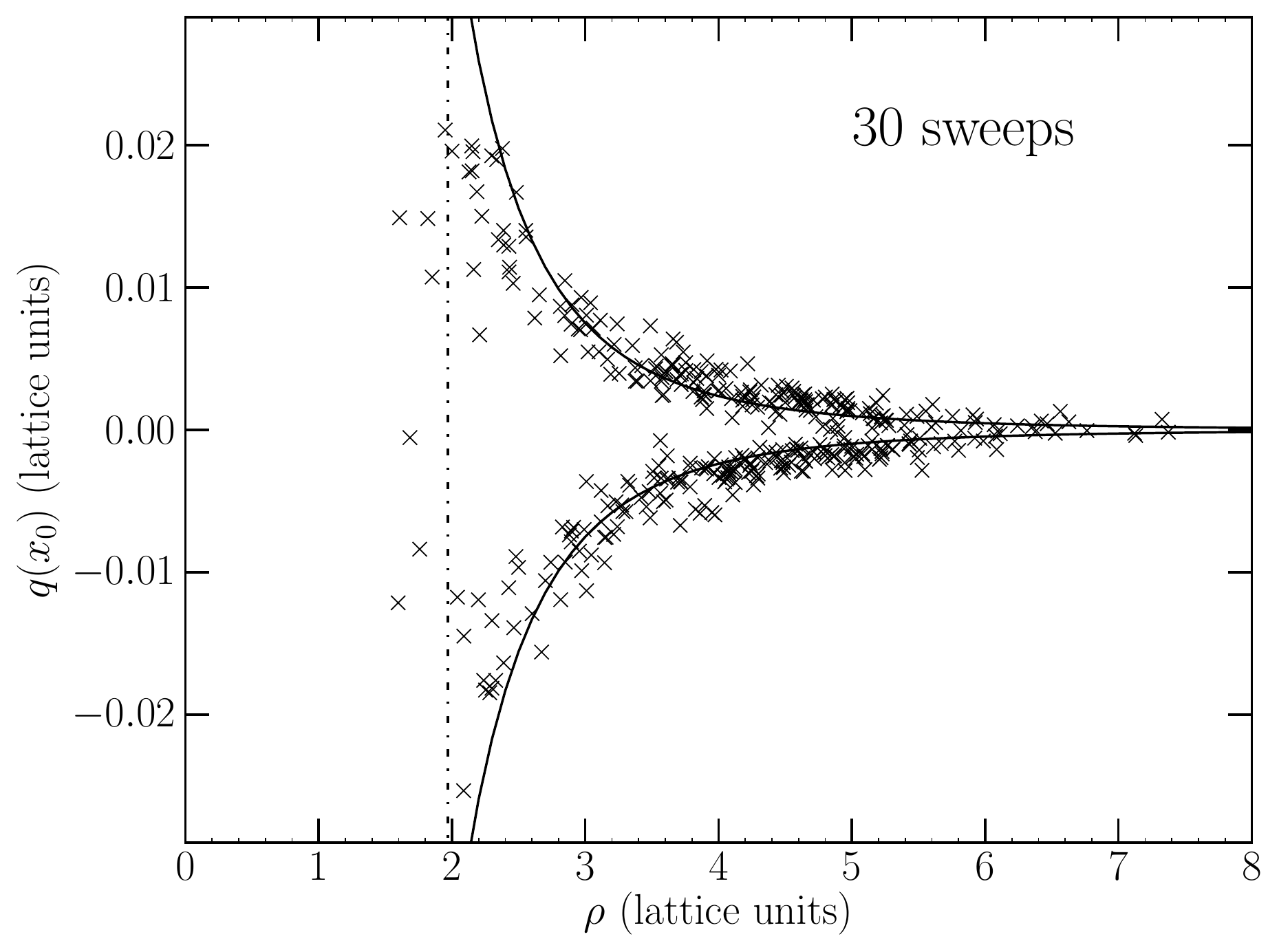}
\includegraphics[ clip=true,width=0.45\hsize]{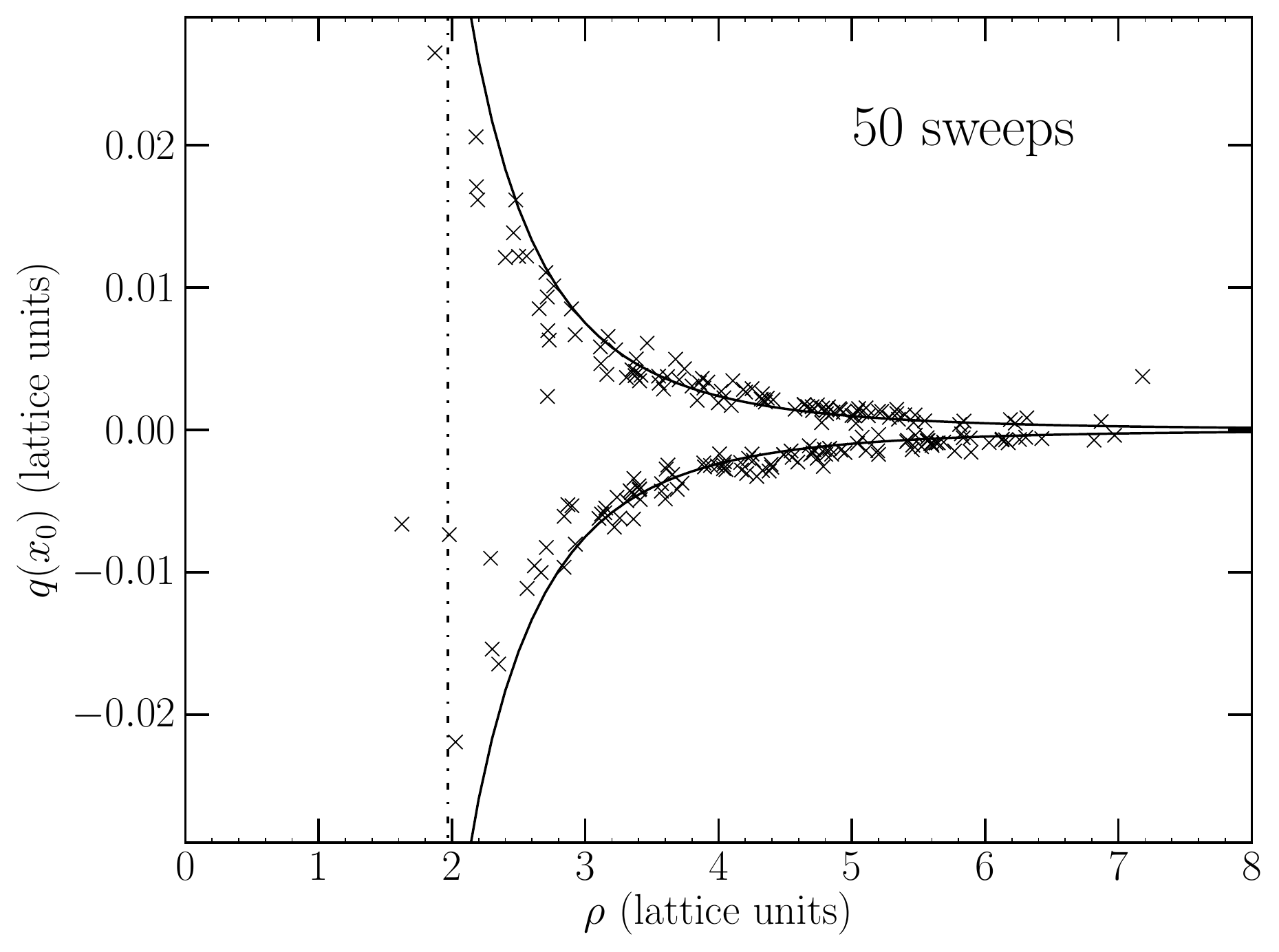}\quad
\includegraphics[ clip=true,width=0.45\hsize]{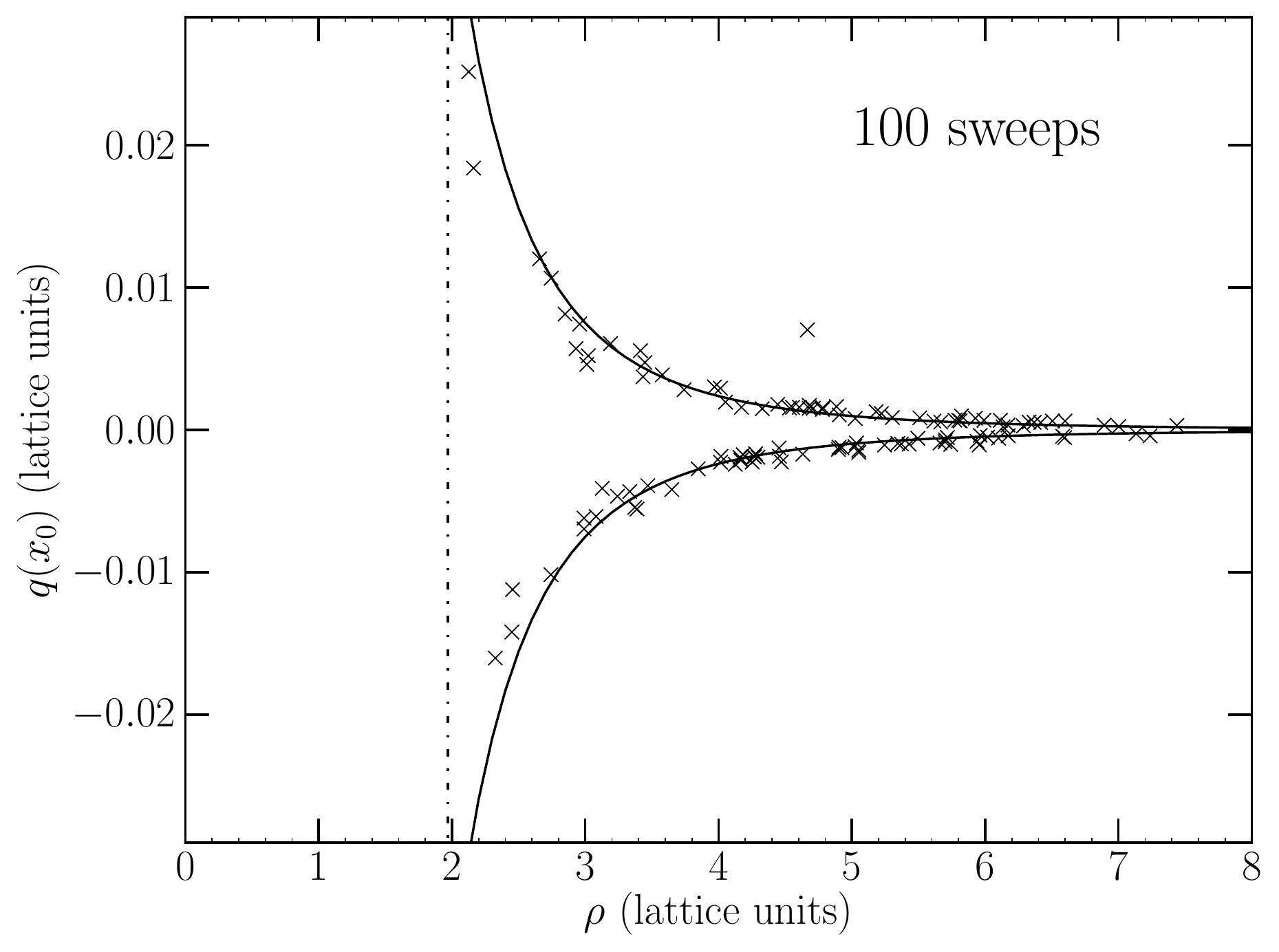}
\end{center}
\caption{Instanton content of a single representative gauge field configuration under the
  over-improved stout-link smearing algorithm at 10, 30, 50 and 100
  sweeps of smearing; reproduced using data from Ref.~\cite{Trewartha:2013qga}.
  At each smearing level, the gauge action $S/S_0 = 2099, 445, 252, 130$; topological charge $Q = -9.45, -7.36, -6.38, -6.02$, and approximate instanton number (determined only by the number of candidates) $n = 1222, 400, 214, 116$.
  The values of the instanton size, $\rho$, found by fitting lattice
  maxima of the action to the classical instanton action density are
  plotted as crosses, against the topological charge at the centre,
  $q(x_{0})$.
  The results are compared to the theoretical relationship between the
  instanton radius and topological charge at the centre (solid lines),
  and the dislocation threshold of the algorithm, $1.97\, a$
  (dash-dotted line). The pseudoparticle densities underlying the gauge fields considered remain higher than the phenomenologically assumed value of $\approx 1~ \rm{fm}^{-4}$, which would correspond to an instanton number of $\approx 80$.}
\label{fig:trewartha}
\end{figure*}

Fig.~\ref{fig:trewartha}, reproduced from
Ref.~\cite{Trewartha:2013qga}, displays the values found on a
smoothed lattice at various smearing levels and compares the
distribution with the classical relationship of Eq.~(\ref{topcharge}),
illustrated by the curves.  The degree to which fitted results concur
with Eq.~(\ref{topcharge}) provides insight into the extent to which
the topology of the gluon fields resembles an ensemble of instantons.

At low levels of smearing we expect to fit a large number of local
maxima which are not associated with the nontrivial topology of
instantons; local maxima of the action corresponding to noise.  At 10
sweeps, the number of instanton candidates is large and distributed
with sizes greater than the dislocation threshold of $\sim 2$ through
8 lattice units.  There is little correlation to the predicted charge
lines of Eq.~(\ref{topcharge}).  However, this quickly changes as the
number of smearing sweeps increases, eventually leading to a very
close approximation to Eq.~(\ref{topcharge}).  In this case the
distribution reflects an ensemble of topological objects approximating
classical instanton solutions. In principle, there are also classical
solutions with non-trivial holonomy (i.e. calorons) which wrap around the temporal
extent of the lattice. In these cases we would not expect the same relationship between
the topological charge and action densities. Such objects would be revealed by an unusually poor or nonconvergent fit to the action density (Eq.~\ref{eq:instactdens}) or by lying far from the expected relationship curve (Eq.~\ref{topcharge}, Fig.~\ref{fig:trewartha}.)
The number of instanton candidates steadily decreases with smearing
through the process of neighboring instanton--anti-instanton
annihilation.  By 100 sweeps the number of instanton candidates has
been thinned to the point that they are usually well-separated and thus
the annihilation of instanton pairs is very slow.  We examine the region
of interest where one has an ensemble of overlapping instantons and
bracket this regime with configuration ensembles having 20, 40 and 60
sweeps of smearing. Using the number of local maxima found in the action density, 
we find pseudoparticle densities of 7.9, 3.6, and 2.4 $\rm{fm}^{-4}$ for 20, 40, and 60 sweep ensembles respectively.

\subsection{Gell-Mann-Oakes-Renner Relation}

We commence with an investigation of the pion mass and consider the
standard pseudoscalar interpolating field $\chi = \bar{u}^a(x)\,
\gamma_5\, d^a(x)$.  Our first consideration is the extent to which
the Gell-Mann-Oakes-Renner \cite{GmOR} relationship between the
squared pion mass and the quark mass ($m_u=m_d=m_q$)
\begin{equation}
m_{\pi}^2 = - \frac{2\, \langle q \bar{q} \rangle}{f_\pi^2}\, m_q \, ,
\end{equation}
is maintained on the smeared ensembles.  We consider a wide range of
hopping parameter values, $\kappa$, and examine the relationship
between the squared pion mass and $1/\kappa \propto m_q$.  

Unlike the case of centre-vortex removal \cite{Elyse}, we find a
linear relationship at all levels of smearing.  This enables the
standard approach of dealing with the additive renormalisation of the
quark mass in Wilson-like fermion formulations such as the FLIC
fermion action considered herein.  The critical value of the hopping
parameter, $\kappa_{\rm cr}$, where the pion mass vanishes is determined
by linearly extrapolating $m_\pi^2$ as a function of $1/\kappa$ to
zero.  The additively renormalised quark mass is then provided by the
standard relation
\begin{equation}
m_q = \frac{1}{2a} \left ( \frac{1}{\kappa} - \frac{1}{\kappa_{\rm cr}} \right ) \, .
\label{qmass}
\end{equation}
The value of $\kappa_{\rm cr}$ observed depends significantly on the
first few sweeps of smearing.  Its deviation from the tree level
value of $1/8$ is an indication of the additive renormalisation of the
quark mass induced by the explicitly broken chiral symmetry of the
Wilson action.  As smearing removes the perturbative physics which
acts to renormalise $\kappa_{\rm cr}$ away from its tree level value,
one observes a return of $\kappa_{\rm cr}$ to 0.125.  Similarly, the
mean link, provided by the fourth root of the average plaquette,
approaches 1.  For example, after 10 sweeps of smearing, the original
value of $\kappa_{\rm cr} = 0.135$ becomes 0.126 and $u_0 = 0.86$
subsequently exceeds 0.995.  The latter result indicates that the
multiplicative renormalisation of the quark mass in the smeared
ensembles is negligible.

\begin{figure}
\begin{center}
\includegraphics[width=1.06\hsize]{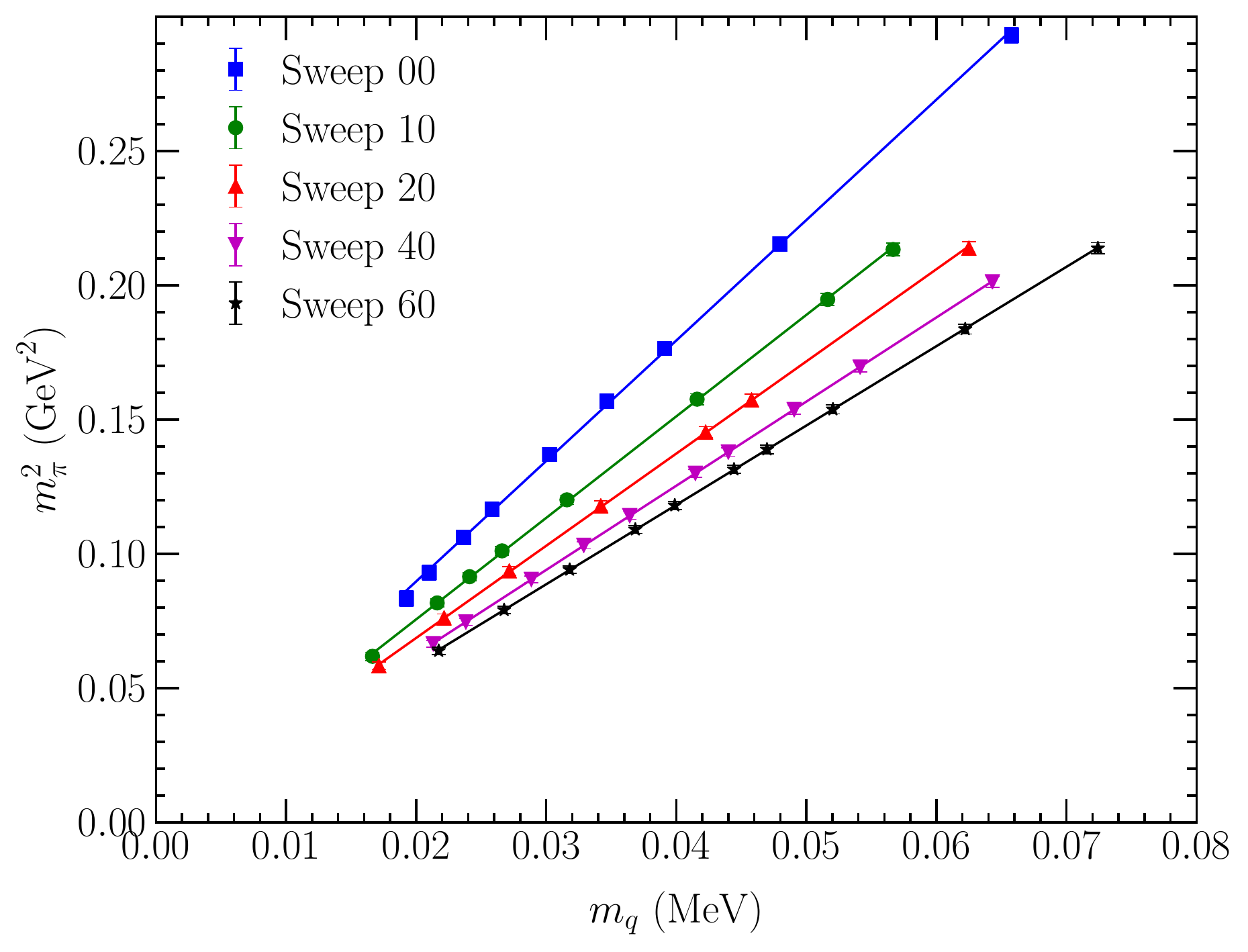}
\end{center}
\caption{The squared pion mass is plotted as a function of the Wilson
  quark mass of Eq.~(\ref{qmass}).  Results are provided for the
  original configurations (Sweep 0) and for the ensembles following
  various levels of over-improved stout-link smearing as indicated.}
\label{fig:gellmann-oakes-renner}
\end{figure}

Our results are illustrated in Fig.~\ref{fig:gellmann-oakes-renner}.
The Goldstone nature of the pion is indeed retained.  However there is
a significant variation in the slope of the linear relation that
reflects an important change in the manner in which the quark mass
manifests itself in the interacting field theory.

\begin{figure}
\begin{center}
\includegraphics[width=0.95\hsize]{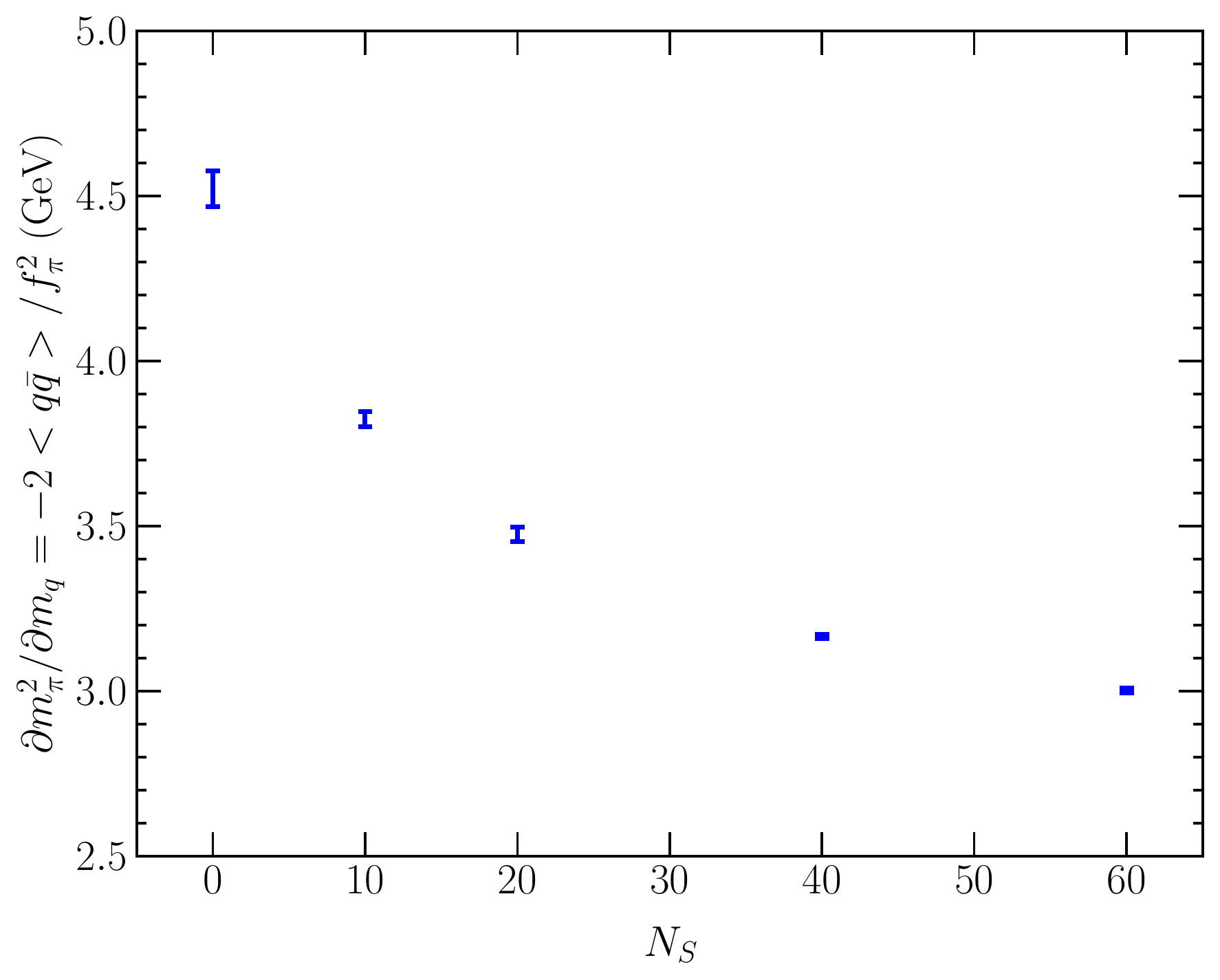}
\end{center}
\caption{The slope of the Gell-Mann-Oakes-Renner relation as a
  function of smearing.}
\label{fig:gmor-coeff}
\end{figure}

The slope, $-2\, \langle q \bar{q} \rangle / f_\pi^2$, can be regarded
as an indicator of the level of instanton preservation under smearing.
Figure~\ref{fig:gmor-coeff} illustrates its evolution under
smearing.  In an instanton model, the quark condensate is proportional to
$\sqrt{n}$, where $n$ is the instanton density.  Thus it is
understandable that the slope reduces under smearing as
instanton--anti-instanton pairs are annihilated.

In comparing results from different levels of smearing, one can choose
to keep the bare mass, $m_q$, fixed, or keep $m_\pi$ fixed as a
measure of the renormalised quark mass.  We choose the latter as
providing the more physical relationship between ensembles with
differing levels of smearing.  It also enables a connection to the
original unsmeared ensemble results.

\subsection{Ground-state hadrons}

We now consider the remaining lowest-lying light hadrons with
non-vanishing masses in the chiral limit: the rho meson, the nucleon,
and the Delta baryon. We use the standard interpolating fields for the $\rho$:
$\bar{u}^a\, \gamma_\mu\, d^a$, nucleon:
$\epsilon^{abc}\, (u^{Ta}\, C \gamma_5\, d^b )\, u^c$, and
$\Delta$: $\epsilon^{abc}\, ( u^{Ta}\, C\gamma_\mu\, u^b)\,
u^c$, and construct a $4 \times 4$ correlation matrix. A 
combination of Gaussians with $N_{sm}=25,50,100,150$ was chosen to
isolate states on the less-smeared configurations ($N_s < 40$). For
the more heavily smeared configurations, the smoother background causes the process
of source smearing to be more efficient - the same number of smearing sweeps gives a larger
source. Thus, the $N_{sm} = 100$ and $N_{sm} = 150 $ sources became too similar in shape
for any linear combination of them to resolve different states. Thus, for these configurations ($N_s=40, N_s=60$), the basis $N_{sm} = 10,25,50,100$ was chosen.

\begin{figure}
\includegraphics[width=0.95\hsize]{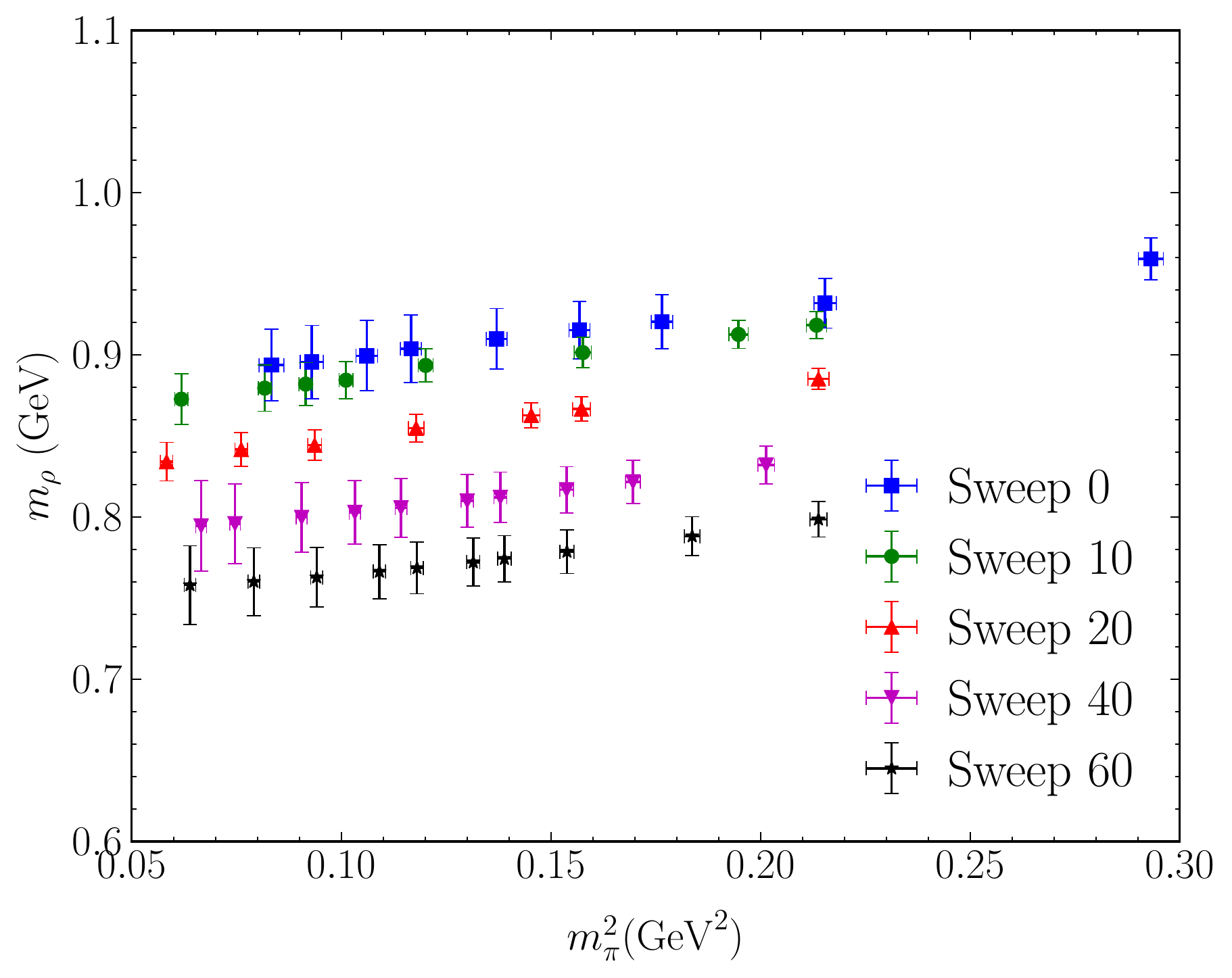}[ht]
\vspace{-7pt}
\caption{Quark-mass dependence of the rho meson mass for various
  levels of smearing.
}
\label{fig:rho-massdep}

\includegraphics[width=0.95\hsize]{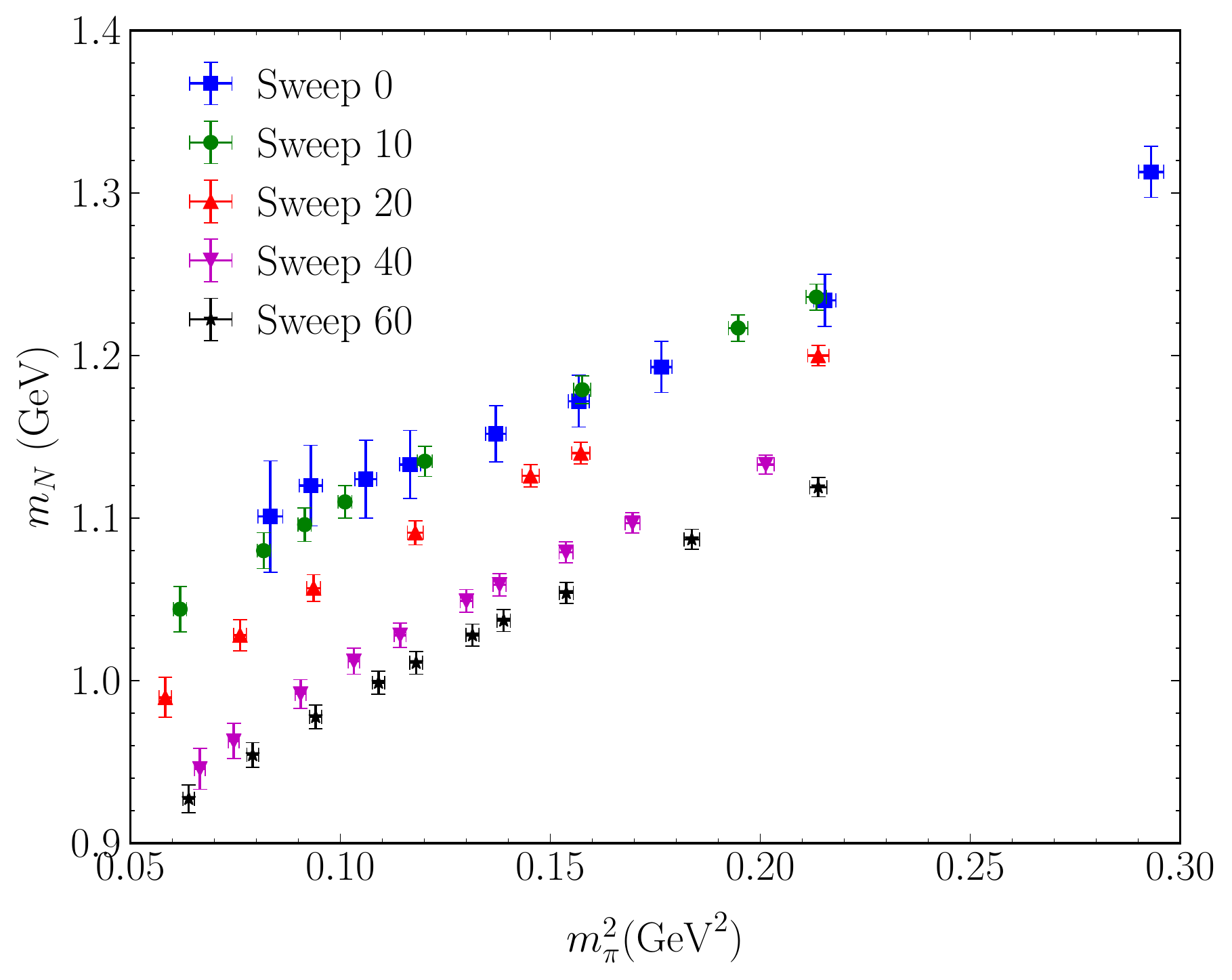}[h]
\vspace{-7pt}
\caption{Quark-mass dependence of the nucleon mass for various levels
  of smearing.
}
\label{fig:nucleon-massdep}

\includegraphics[width=0.95\hsize]{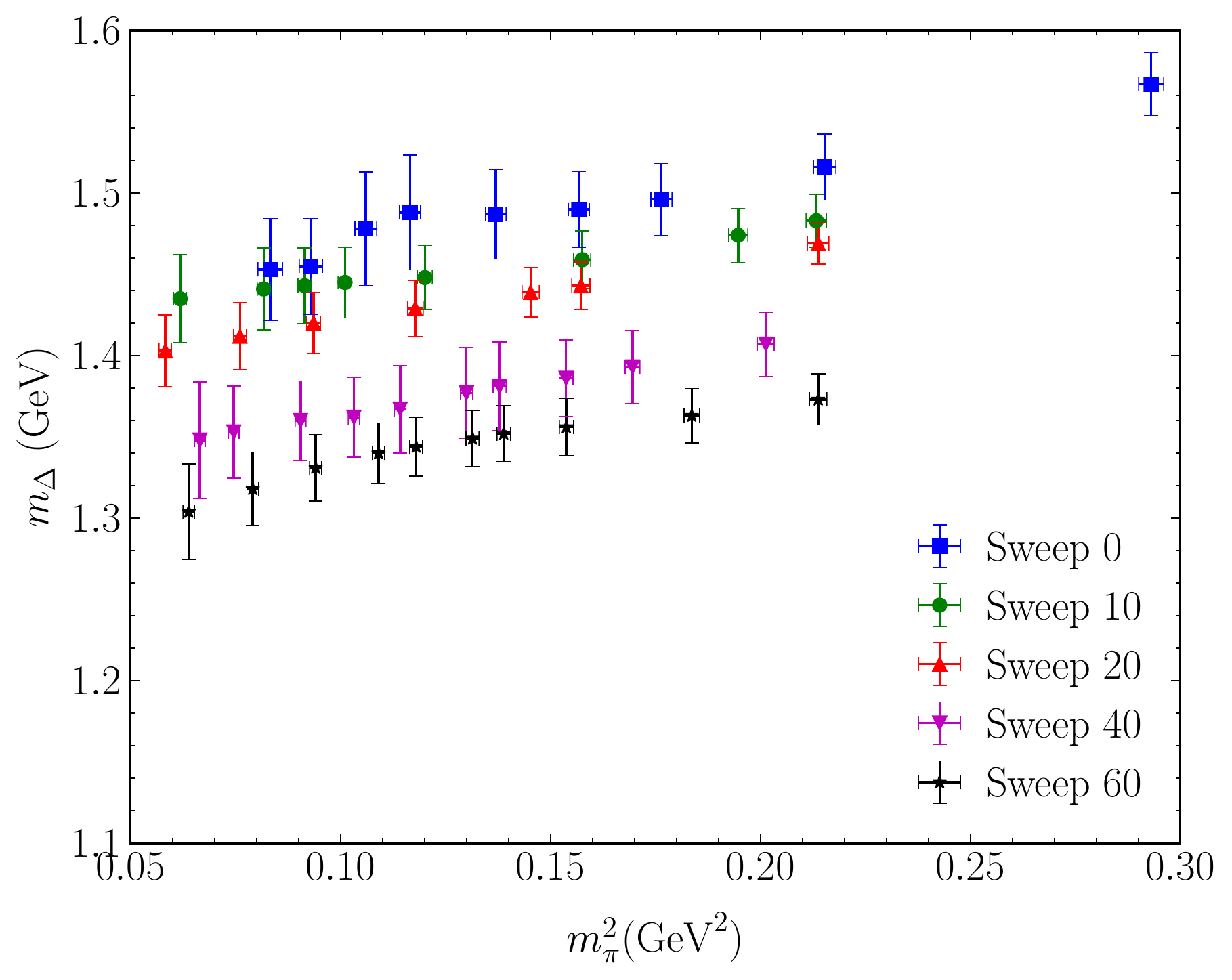}[h]
\vspace{-7pt}
\caption{Quark-mass dependence of the $\Delta$ mass for various levels
  of smearing.
}
\label{fig:delta-massdep}
\end{figure}

As illustrated in Figs.~\ref{fig:rho-massdep} through
\ref{fig:delta-massdep}, the hadron masses display a common trend of
reduction as the underlying instanton content of the vacuum is eroded
through pair annihilation under smearing.  Only subtle changes in the
pion mass dependence of the hadron masses are observed.

However, an important difference between these hadronic observables
and gluonic observables is apparent.  While gluonic observables such
as the action, and the short-distance potential undergo rapid
transitions during the first few sweeps of smearing, these hadronic
observables display very little change over the first 10 sweeps.  One
can conclude that the rapid loss of action density in the first few
sweeps of cooling is not connected to the low-lying hadron masses in a
significant manner.  Rather it is the loss of closely spaced
instanton-anti-instanton pairs over more extensive smearing extents
that gives rise to a loss of dynamical mass generation and lower lying
hadron masses.

However, we emphasize the pion is different.  At larger values of the
quark mass, the pion mass displays a more rapid transition over the
first few sweeps as reported in Fig.~\ref{fig:gellmann-oakes-renner}.
This behaviour contrasts the $\rho$, $N$ and $\Delta$ where the
change in the hadron mass is relatively insensitive to the quark mass.

To perform a more quantitative examination of these hadron masses we
interpolate the results to a common pion mass of 300 MeV and increase
the statistical sample size by calculating 8 fermionic sources (shifting by
a quarter of the temporal extent, and by half the spatial extent in each
direction) per configuration for a total of 608.  We also consider an interpolation
to $m_\pi = 400$ MeV to expose any sensitivity to our selection of a comparison point.  We note that
for $m_\pi = 300$ MeV, $m_\pi L \sim 4$ such that finite-size effects
are unlikely to affect the results in a significant manner.

\begin{figure}
\includegraphics[width=0.95\hsize]{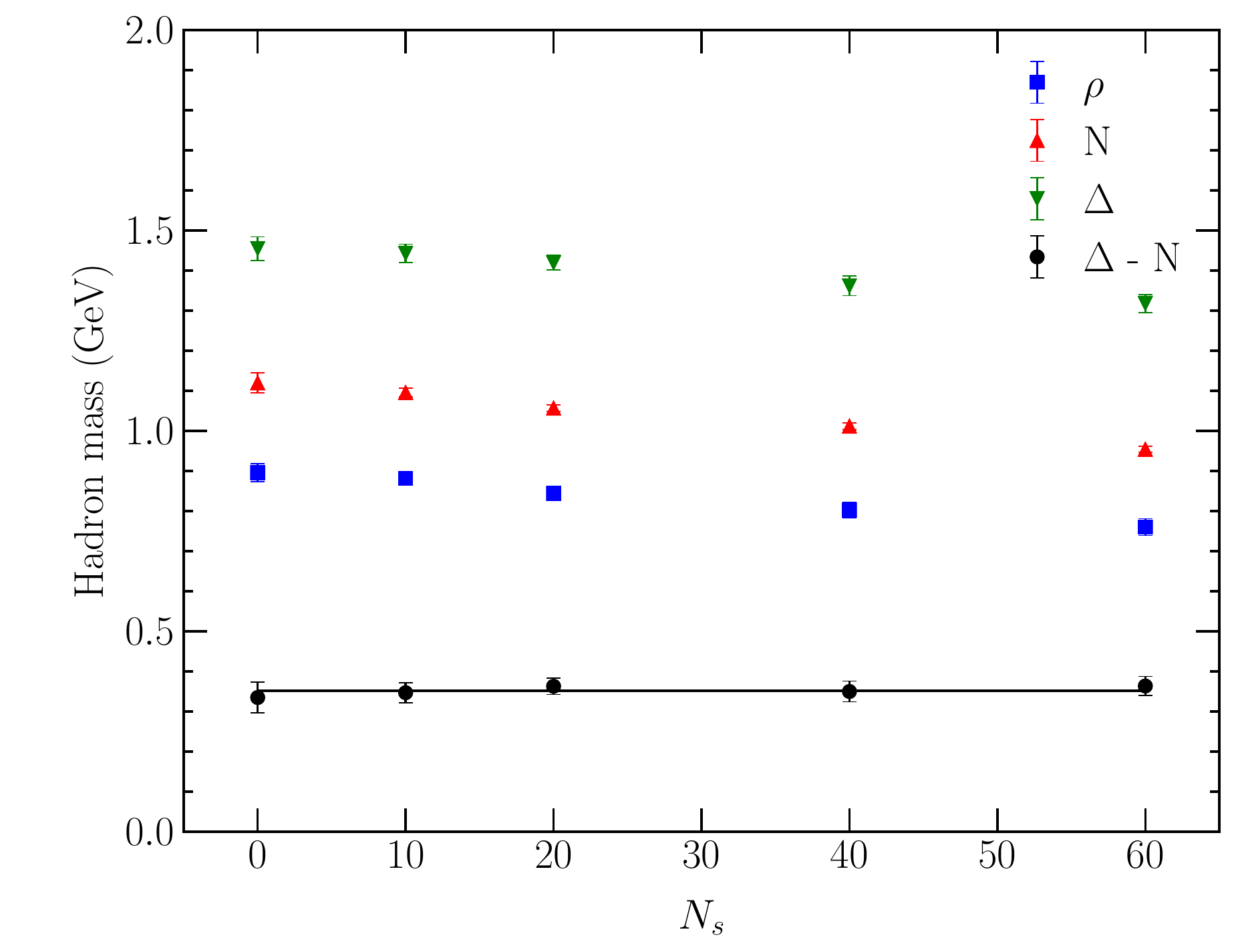}
\caption{Hadron masses, interpolated to a common pion mass of 300 MeV,
  are illustrated as a function of the number of smearing sweeps,
  $N_s$.  The nucleon-Delta mass splitting, $M_\Delta - M_N$, is also
  illustrated.  A fit of this splitting to a constant illustrates the
  invariance of the nucleon-Delta mass splitting to a thinning of the
  (anti-)instanton density.}  
\label{fig:hadrons1}
\end{figure}

\begin{figure}
\includegraphics[width=0.95\hsize]{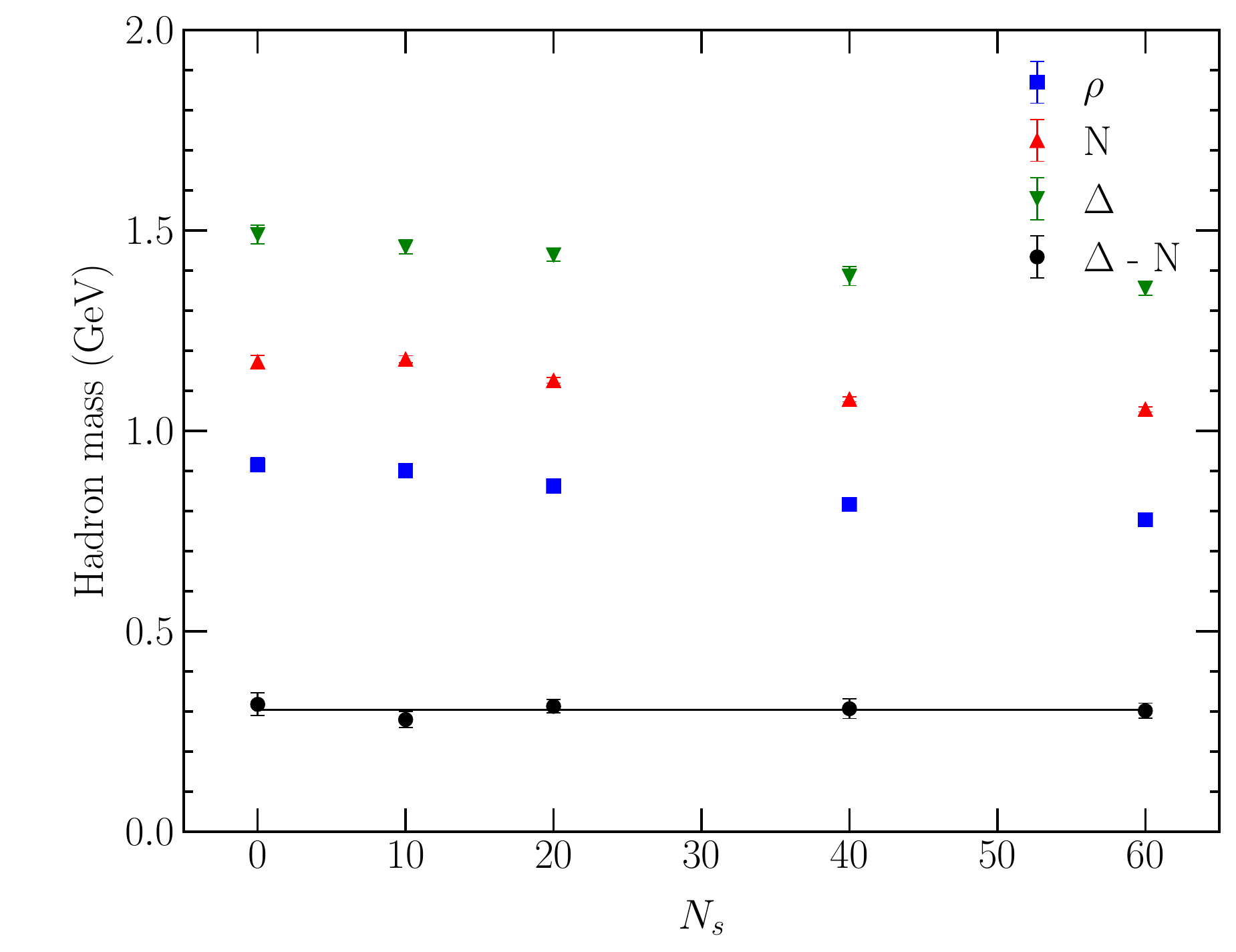}
\caption{Hadron masses, interpolated to a common pion mass of 400 MeV,
  are illustrated as a function of the number of smearing sweeps,
  $N_s$.  A fit of the nucleon-Delta mass splitting, $M_\Delta - M_N$,
  to a constant illustrates the invariance of the nucleon-Delta mass
  splitting to a thinning of the (anti-)instanton density.}
\label{fig:hadrons2}
\end{figure}

Figures~\ref{fig:hadrons1} and \ref{fig:hadrons2} report the results.
While the hadron masses show a decline under smearing, we note that
this decline is uniform for the $N$ and $\Delta$.  The mass splitting
$M_\Delta - M_N$ is invariant under smearing.

This invariance is interesting in the context of an instanton model where a strong attractive contribution to the nucleon mass originates from the interaction of the scalar diquark component with a single instanton or with a pair of an instanton and an anti-instanton (in the sum-rule context, the single-instanton contribution is necessary to stabilise the correlator) \cite{Shuryak-prop-3, ZPhysC.46.281, PhysRevLett.71.484}. This contribution is necessarily large in typical models in order to replicate the observed nucleon-Delta mass splitting. In contrast, the lowest order contribution to the Delta requires twice as many zero-mode contributions and the instanton contribution to the mass is therefore higher order in $n/V$. In the current investigation, once the ultraviolet interactions have been suppressed under smearing, the main change to the gauge fields under further smearing is the annihilation of adjacent instanton-anti-instanton pairs.  As the instanton-based model experiences a reduction in the scalar diquark attraction in the nucleon, it predicts a narrowing of the nucleon-Delta mass splitting under smearing. However, this is not observed.  Figures~\ref{fig:hadrons1} and
\ref{fig:hadrons2} display a nucleon-Delta mass splitting that is
invariant under smearing.  Thus, the simple direct-instanton effect
cannot be responsible for the lightness of the nucleon compared to the
Delta.  

In summary, the nucleon and Delta masses decrease under over-improved
stout-link smearing.  The reduction is the order of 10\% of their
original masses after 60 sweeps of smearing.  However the mass
splitting between them is insensitive to the loss of instanton pairs
under smearing.  Their mass splitting remains constant within error.
The loss of instanton-induced scalar-diquark attraction in the nucleon
predicted by the simple direct-instanton effect is not apparent.
Although the majority of the hadron mass can be considered as
generated by instanton interactions, this hadron-instanton interaction
is not described using the `tHooft interaction from a single instanton
pair.

It is important to place this modern analysis in the context of an
early analysis \cite{Chu:1994vi} that reported results consistent with
the instanton model prediction.  This early study considered smaller
$16^3 \times 24$ lattices with an uncooled lattice spacing of 0.168 fm.
The ensemble consisted only of 19 gauge configurations, and the hadron masses on the cooled ensembles were fitted using their dispersion relation because the statistical variation was too large on the cooled configurations to use the asymptotic behaviour of $e^{-mt}$.
The associated large discretisation errors combined with unimproved
cooling led to the rapid reduction of both ultraviolet physics and
instanton content.  Individual instantons are destroyed under
unimproved cooling leading to a rapid loss of gauge field dynamics.
Using the nucleon mass to reset the lattice spacing after cooling, the
Delta was found to be ``too light.''  The present analysis suggests
that this loss of mass splitting reflects a loss of gauge field
dynamics such that the nucleon and Delta simply become degenerate.  In
other words, the dynamics responsible for splitting them has been
destroyed under unimproved cooling.  

\begin{figure}
\includegraphics[width=0.95\hsize]{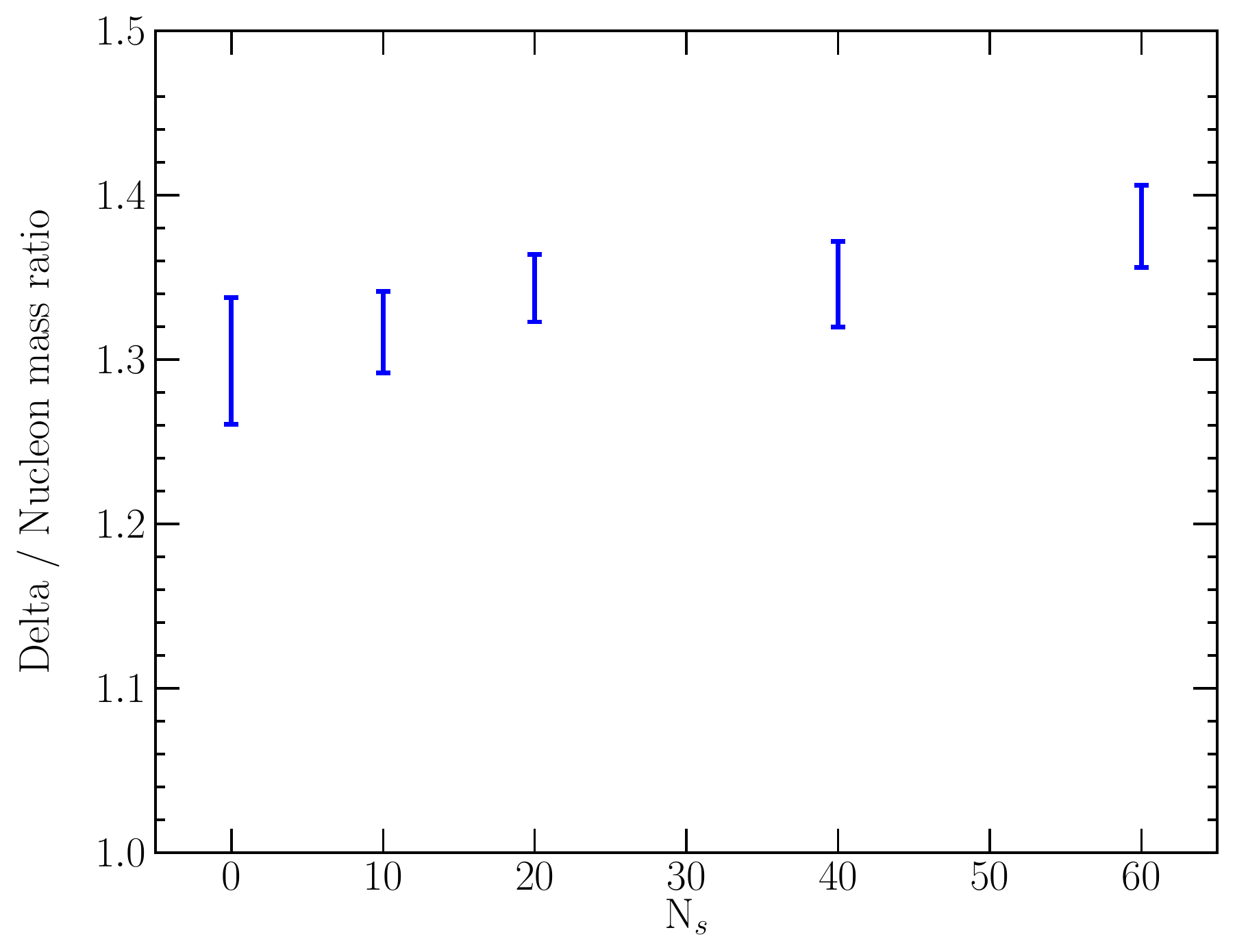}
\caption{The dependence of the $\Delta/N$ mass ratio evaluated at
  $m_\pi = 300$ MeV on the number of smearing sweeps.  The ratio
  increases with smearing, reflecting a constant mass splitting in the
  context of decreasing baryon masses due to a thinning of the
  (anti-)instanton density.}
\label{fig:DeltaNmassRatio}
\end{figure}

In contrast, the invariant mass splitting observed in the present
analysis occurs in the context of declining nucleon and Delta masses.
As a result, the Delta-nucleon mass ratio actually increases under
smearing as illustrated in Fig.~\ref{fig:DeltaNmassRatio}.  While good
agreement with the physical ratio of 1.31 is observed at small numbers
of smearing sweeps, the Delta is ``too heavy'' at 60 sweeps of
smearing when measured relative to the nucleon mass.

\subsection{Excited States}

Our use of the correlation matrix method to cleanly isolate the
lowest-lying states presented thus far has the additional advantage
that we are able to examine the behaviour of excited states under
smearing.  To the best of our knowledge, this is the first examination
of the role of instanton degrees of freedom in describing the radial
excitations of hadrons.

\begin{figure}
\includegraphics[width=0.95\hsize]{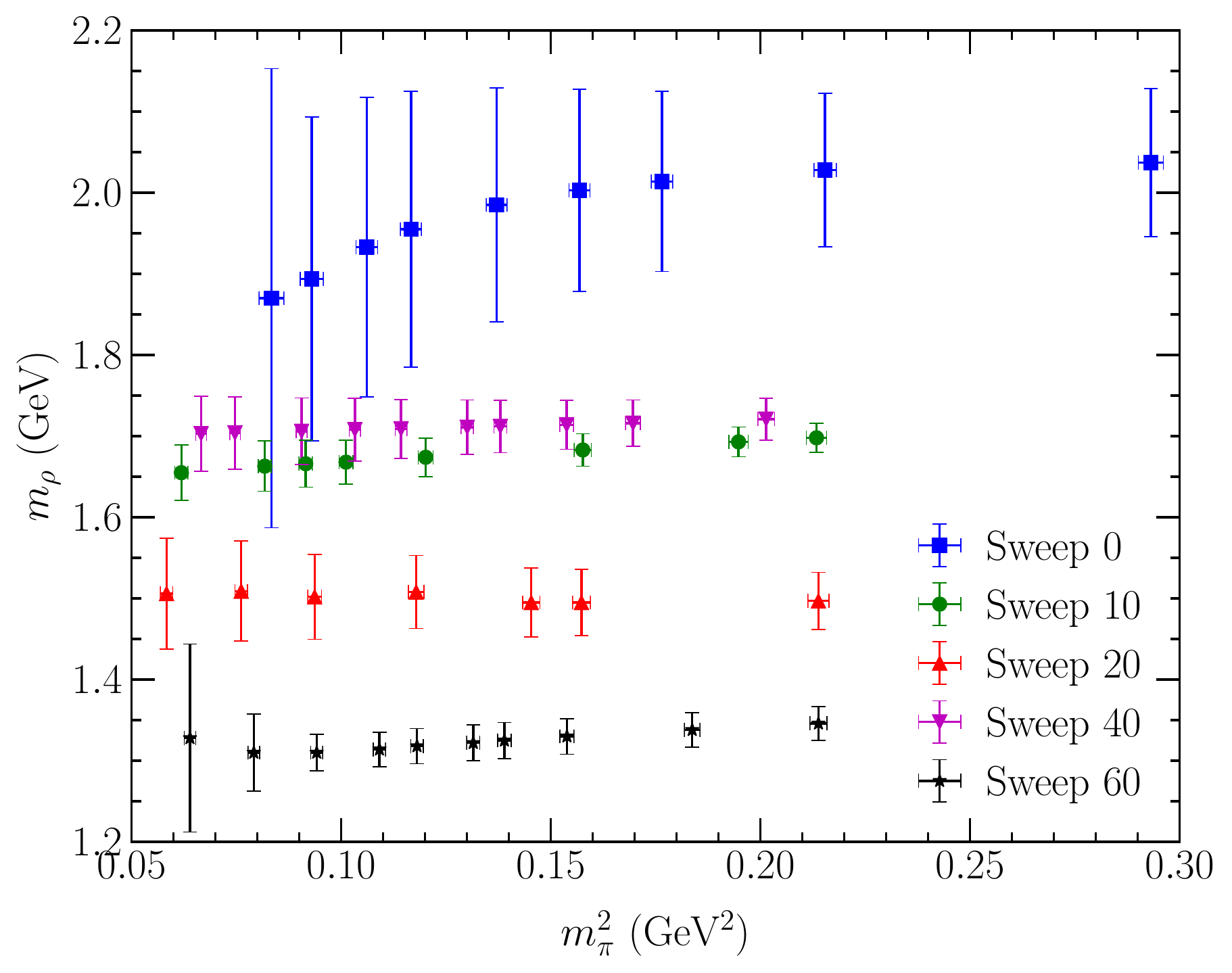}
\vspace{-7pt}
\caption{Quark-mass dependence of the first excited-state energy of
  the rho meson observed in our correlation-matrix analysis for
  various levels of smearing.}
\label{fig:rho-ex}

\includegraphics[width=0.95\hsize]{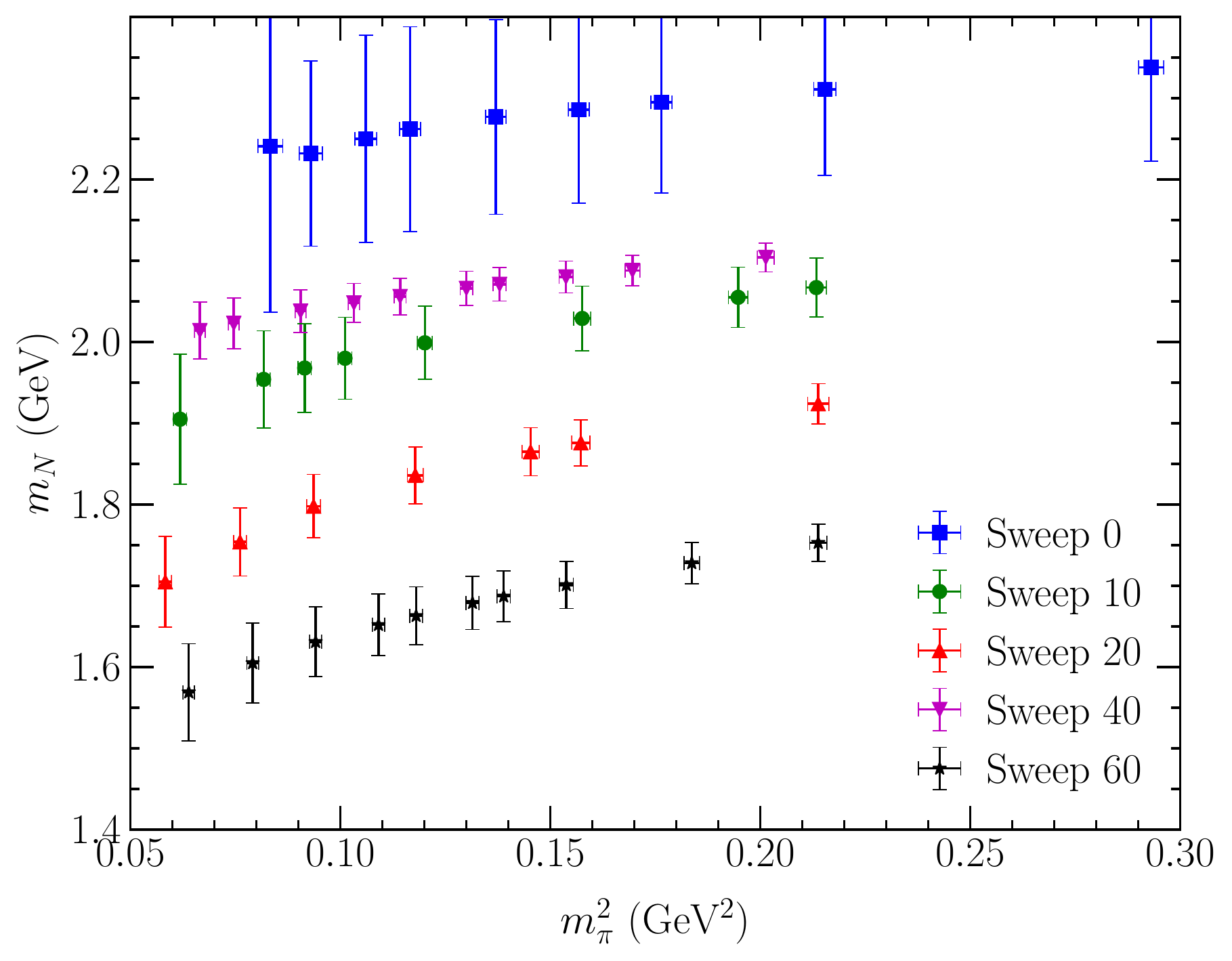}
\vspace{-7pt}
\caption{Quark-mass dependence of the first excited-state energy of
  the nucleon observed in our correlation-matrix analysis for various
  levels of smearing.}
\label{fig:nucleon-ex}

\includegraphics[width=0.95\hsize]{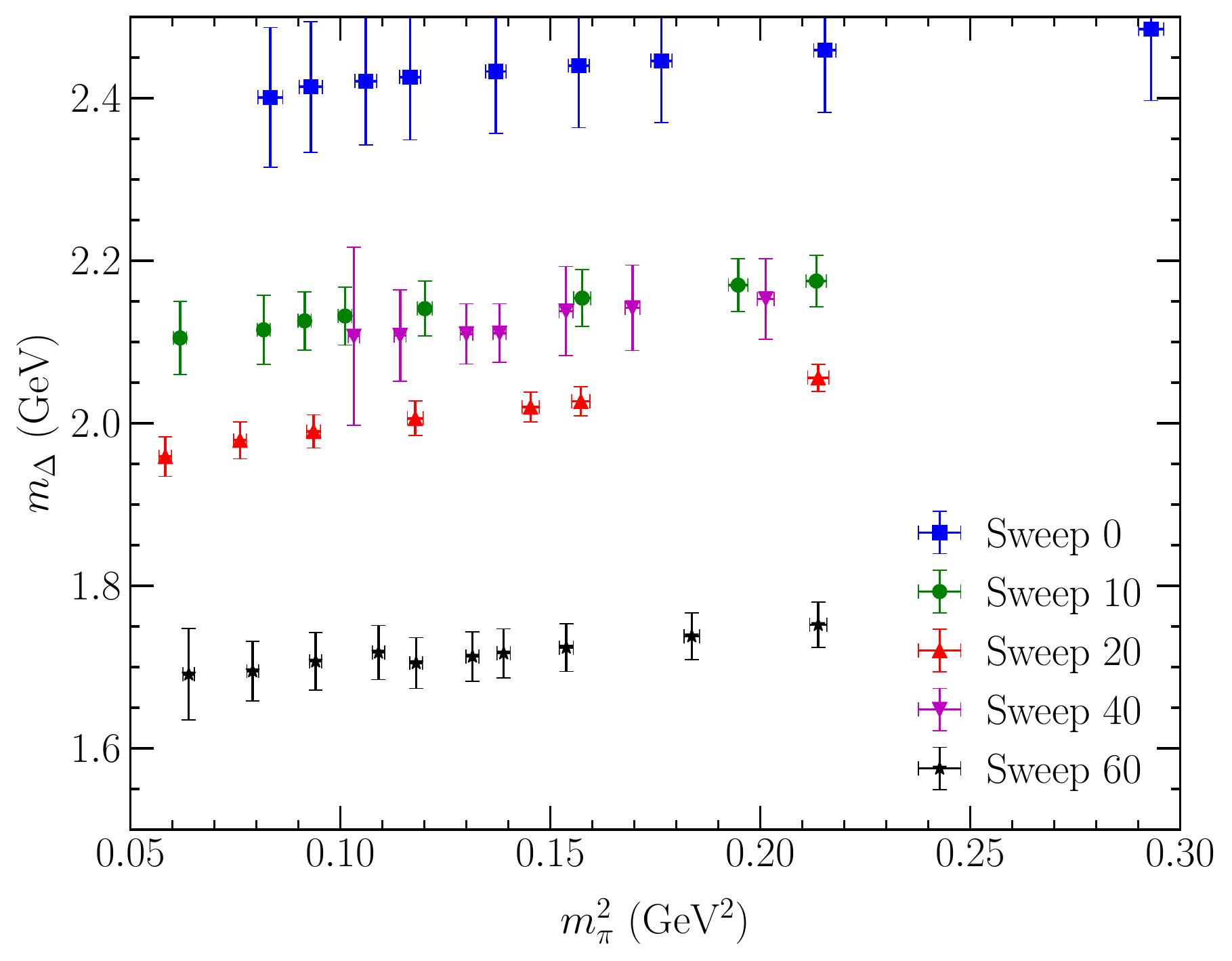}
\vspace{-7pt}
\caption{Quark-mass dependence of the first excited-state energy of
  the Delta observed in our correlation-matrix analysis for various
  levels of smearing.  }
\label{fig:delta-ex}
\end{figure}

Figures \ref{fig:rho-ex} through \ref{fig:delta-ex} illustrate the
very different behaviour of hadronic excitations under smearing.  The
variation in mass is not at the level of the 10\% observed for ground
states, but is the order of 30\%.  Moreover, a significant mass drop
is observed for as few as 10 sweeps of smearing, signaling an
important role for ultraviolet physics.  

We note that the excited states at 40 sweeps of smearing do not follow
the monotonic trend set by the other smearing levels.  We believe
these energies are affected by near-by states that are not adequately
accommodated in the $4 \times 4$ correlation matrix considered, giving
rise to a superposition of excited states in the reported results.
While we include them here for completeness, we will set them aside
for the remainder of the discussion.

\begin{figure}
\begin{center}
\includegraphics[width=\hsize]{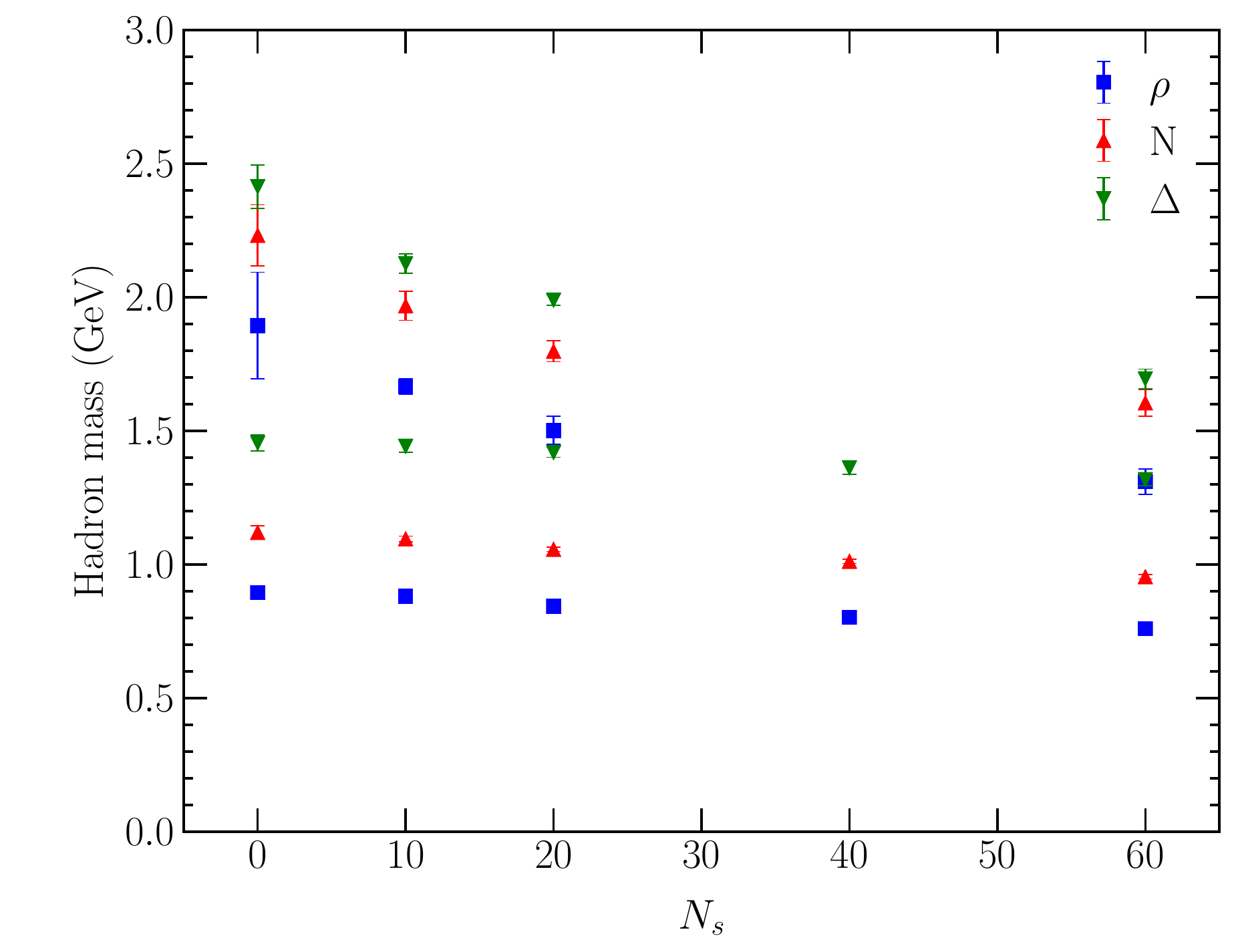}
\end{center}
\caption{The observed energies of the first excited states of our
  correlation matrix analysis are interpolated to a common pion mass
  of 300 MeV and plotted as a function of the number of smearing sweeps,
  $N_s$, to reveal the role of instanton degrees of freedom in
  generating the spectrum of excited states. The ground state masses are also replotted for comparison. }
\label{fig:exc1}
\end{figure}

\begin{figure}
\begin{center}
\includegraphics[width=\hsize]{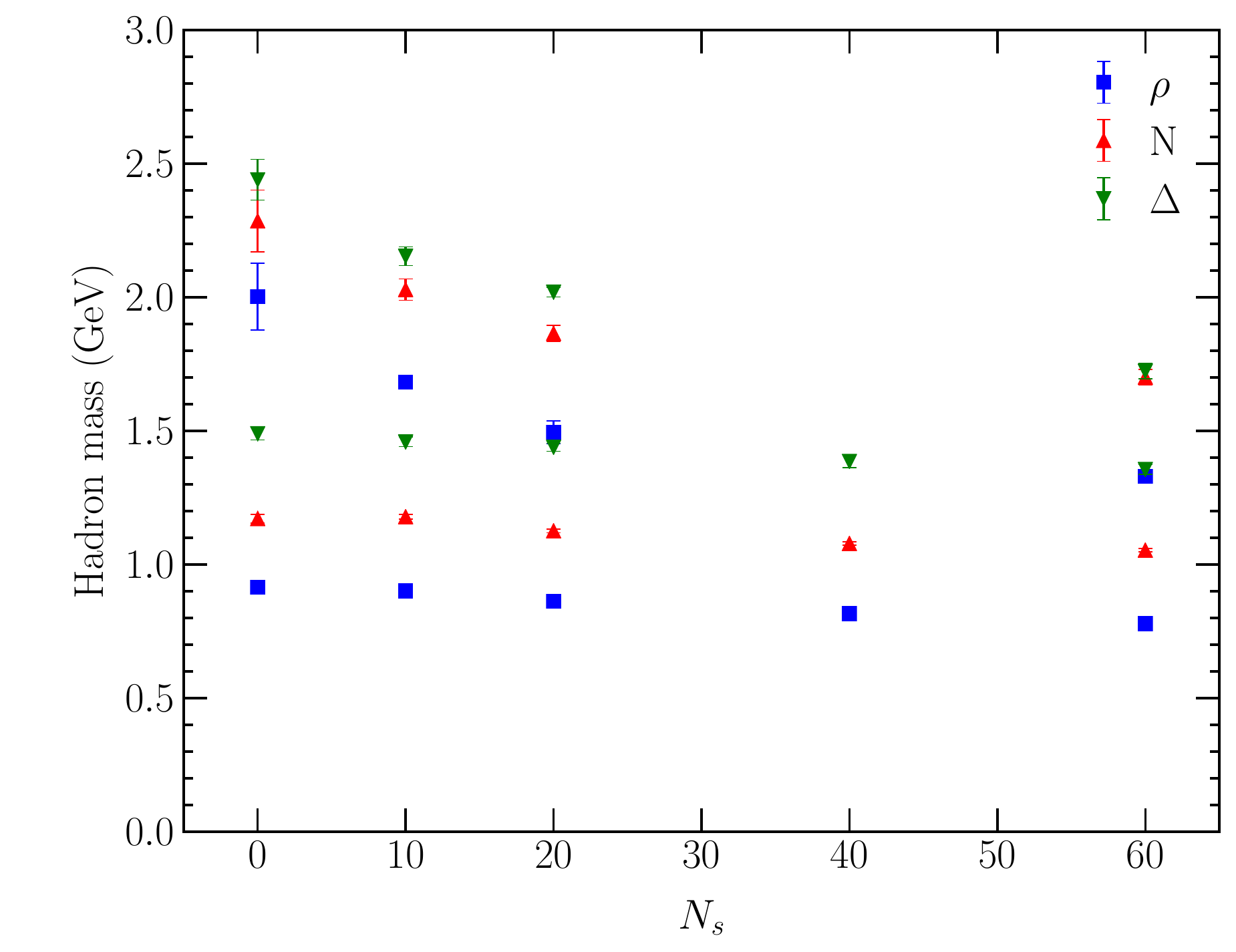}
\end{center}
\caption{The observed energies of the first excited states of our
 correlation matrix analysis are interpolated to a common pion mass
  of 400 MeV and plotted as a function of the number of smearing sweeps,
  $N_s$, to reveal the role of instanton degrees of freedom in
  generating the spectrum of excited states. Again, the ground state masses are shown for comparison.}
\label{fig:exc2}
\end{figure}

As for the ground states, we interpolate the hadron masses to a common
pion mass of 300 and 400 MeV.  Their dependence on the number of
smearing sweeps is illustrated in Figs.~\ref{fig:exc1} and
\ref{fig:exc2}.  While the ground-state hadrons remained qualitatively
unchanged under smearing, the excited states decrease in mass
significantly.  Continued smearing leads to a significant decline in
the excitation energy, again emphasizing an important role for
instanton degrees of freedom in generating the excitation spectrum.

An analysis of the associated eigenvectors for the excited states
indicates that these states have the expected single-node structure of
a radial excitation in their wave function.  The node is generated
through a superposition of broad and narrow Gaussian smeared sources
with opposite signs.


\subsection{Non-relativistic Quark Model Consideration}

An interesting question is the extent to which our results reported in
Figs.~\ref{fig:exc1} and \ref{fig:exc2} can be described by a simple
constituent quark model, drawing on the change of the static quark potential
examined in Fig.~\ref{fig:sqp}.

Here we consider the $\rho$ meson.  Using our fits to the static quark
potential illustrated in Fig.~\ref{fig:sqp}, we solve the
Schr\"{o}dinger equation using a fixed constituent quark mass of 400
MeV and boundary conditions emulating the periodic lattice condition
of the spatial volume.
\begin{figure}
\begin{center}
\includegraphics[width=\hsize]{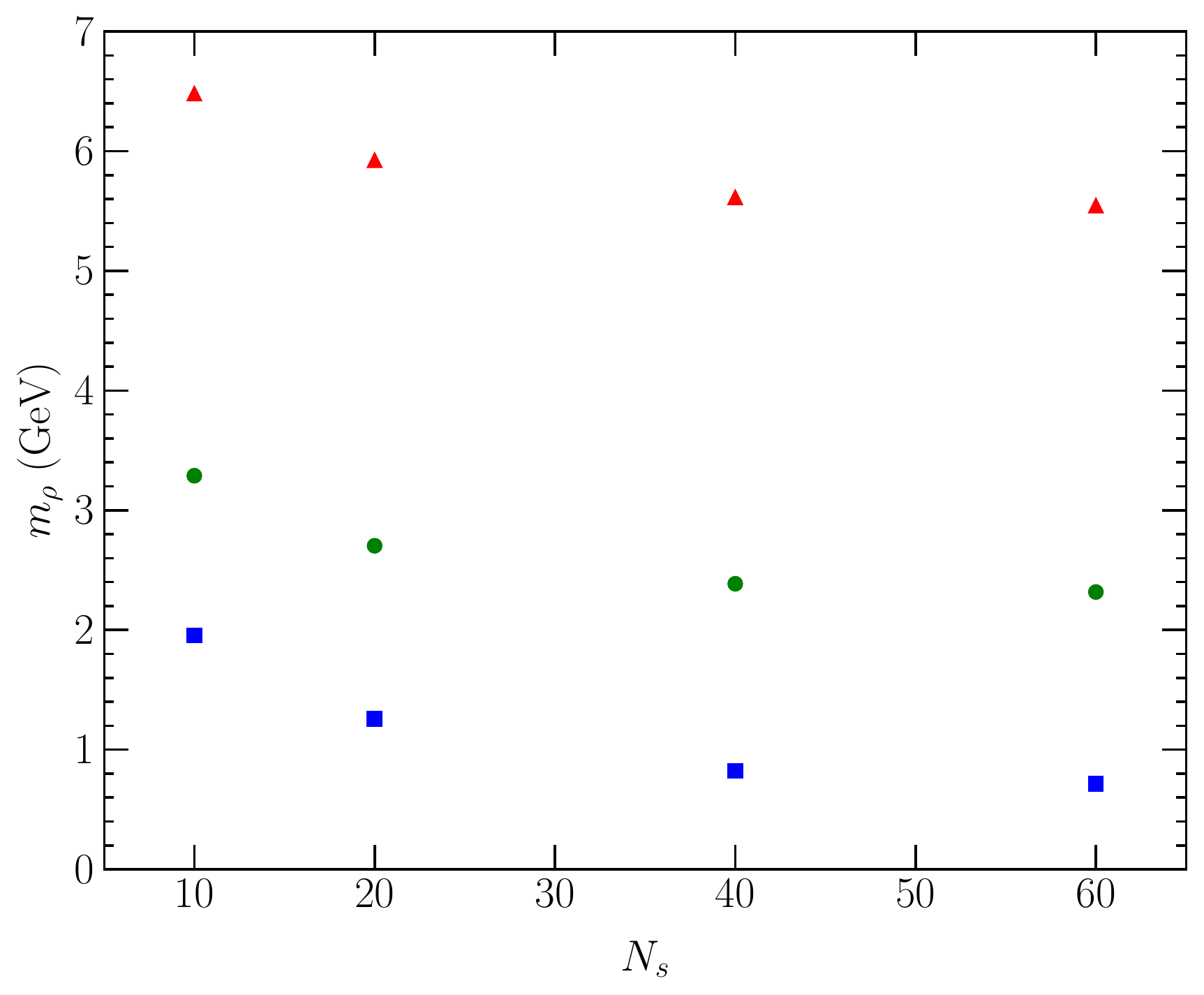}
\end{center}
\caption{Ground, first- and second-excited state masses of the $\rho$
  meson from a nonrelativistic quark-model calculation based on the
  static quark potential fits illustrated in Fig.~\ref{fig:sqp}.
  Masses may be shifted vertically by introducing a constant into the
  potential.}
\label{fig:nrqm}
\end{figure}

As displayed in Fig.~\ref{fig:nrqm}, this naive model predicts a much
faster decrease in the ground-state mass than we observe in the
lattice calculation.  Moreover, the excited state mass is maintained
better than the ground state under smearing.  This qualitative
difference allows us to conclude that a simple modification of the
potential energy between constituent quarks is insufficient to capture
the essence of the modification of the QCD vacuum under smearing.

\section{Concluding Remarks}
\label{sec:conclusion}

The light hadron spectrum has been examined in lattice QCD where the
vacuum is altered using the over-improved stout-link smearing
algorithm designed to retain separated instantons. The change in the
ground-state hadron masses of the $\rho$, $N$ and $\Delta$ is of the
order of 10\%, indicating that almost all of the mass is generated by
topological structures similar to instantons. 

However, the difference between the Delta and nucleon masses is insensitive to smearing and the associated thinning of (anti-)instantons on the lattice.  Even though the smearing process destroys topology by pairwise annihilation, the anticipated attractive contribution to the nucleon mass from the scalar-diquark direct-instanton interaction does not weaken during this process. This indicates that direct instanton-induced effects are not the dominant contribution to the hadronic masses.

Similarly, simple quark model phenomenology differs from the results
observed on the lattice.  A deeper understanding of the underlying
mechanisms of QCD could be obtained through the operator product expansion of
two-point hadron correlation functions.  The evolution of vacuum
condensates under smearing could be examined and the impact of this
evolution on the spectral properties of the correlators could be
studied.  We will leave this study for a future investigation.
\\
\section{Acknowledgements}
\label{sec:Acknowledgements}
It is a pleasure to acknowledge James Zanotti, Jon-Ivar Skullerud and
Daniel Trewartha for interesting discussions.  This research was
undertaken on the NCI National Facility in Canberra, Australia, which
is supported by the Australian Commonwealth Government. We also
acknowledge eResearch SA for generous grants of supercomputing
time. This research is supported by the Australian Research Council under grants DP1201104627 and DP15013164.

\bibliography{instanton_2014}
\bibliographystyle{h-physrev}
\end{document}